\documentclass[twocolumn]{aastex63}
\usepackage[utf8]{inputenc}
\usepackage{enumitem}

\usepackage{amsmath}
\usepackage{graphicx}
\usepackage{hyperref}
\usepackage{natbib}
\usepackage{color}
\usepackage{comment}
\usepackage[normalem]{ulem}
\usepackage{academicons}
\usepackage{rotating} 
\usepackage{lineno}

\definecolor{dukeblue}{rgb}{0.0, 0.0, 0.61}
\hypersetup{colorlinks=True,urlcolor=cyan,citecolor={blue},linkcolor={red}}
\newcommand{\ark}[1]{#1}

\begin{document}

\title[Non-linear Timing]{Generalized \ark{Non-linear} Bayesian Pulsar Timing with \texttt{Enterprise}}
\author[0000-0002-3654-980X]{Andrew R. Kaiser}
\affiliation{Department of Physics and Astronomy, West Virginia University, P.O. Box 6315, Morgantown, WV 26506, USA}
\affiliation{Center for Gravitational Waves and Cosmology, West Virginia University, Chestnut Ridge Research Building, Morgantown, WV 26505, USA}

\author[0000-0003-2742-3321]{Jeffrey S. Hazboun}
\affiliation{Department of Physics, Oregon State University, Corvallis, OR 97331, USA}

\author[0000-0001-7697-7422]{Maura A. McLaughlin}
\affiliation{Department of Physics and Astronomy, West Virginia University, P.O. Box 6315, Morgantown, WV 26506, USA}
\affiliation{Center for Gravitational Waves and Cosmology, West Virginia University, Chestnut Ridge Research Building, Morgantown, WV 26505, USA}

\author[0000-0002-6039-692X]{H. Thankful Cromartie}
\altaffiliation{NASA Hubble Fellowship: Einstein Postdoctoral Fellow}
\affiliation{Cornell Center for Astrophysics and Planetary Science and Department of Astronomy, Cornell University, Ithaca, NY 14853, USA}

\author[0000-0001-8384-5049]{Emmanuel Fonseca}
\affiliation{Department of Physics and Astronomy, West Virginia University, P.O. Box 6315, Morgantown, WV 26506, USA}
\affiliation{Center for Gravitational Waves and Cosmology, West Virginia University, Chestnut Ridge Research Building, Morgantown, WV 26505, USA}

\author[0000-0003-1407-6607]{Joseph Simon}
\altaffiliation{NSF Astronomy and Astrophysics Postdoctoral Fellow}
\affiliation{Department of Astrophysical and Planetary Sciences, University of Colorado, Boulder, CO 80309, USA}

\author[0000-0003-0264-1453]{Stephen R. Taylor}
\affiliation{Department of Physics and Astronomy, Vanderbilt University, 2301 Vanderbilt Place, Nashville, TN 37235, USA}

\author[0000-0002-4162-0033]{Michele Vallisneri}
\affiliation{Jet Propulsion Laboratory, California Institute of Technology, 4800 Oak Grove Drive, Pasadena, CA 91109, USA}
\affiliation{Division of Physics, Mathematics, and Astronomy, California Institute of Technology, Pasadena, CA 91125, USA}

\author[0000-0003-4700-9072]{Sarah J. Vigeland}
\affiliation{Center for Gravitation, Cosmology and Astrophysics, Department of Physics, University of Wisconsin-Milwaukee,\\ P.O. Box 413, Milwaukee, WI 53201, USA}

\author{Zaven Arzoumanian}
\affiliation{X-Ray Astrophysics Laboratory, NASA Goddard Space Flight Center, Code 662, Greenbelt, MD 20771, USA}

\author[0000-0003-2745-753X]{Paul T. Baker}
\affiliation{Department of Physics and Astronomy, Widener University, One University Place, Chester, PA 19013, USA}

\author[0000-0003-4046-884X]{Harsha Blumer}
\affiliation{Department of Physics and Astronomy, West Virginia University, Morgantown, WV 26506-6315, USA}
\affiliation{Center for Gravitational Waves and Cosmology, Chestnut Ridge Research Building, Morgantown, WV 26505, USA}

\author[0000-0003-3053-6538]{Paul R. Brook}
\affiliation{Institute for Gravitational Wave Astronomy and School of Physics and Astronomy, University of Birmingham, Edgbaston, Birmingham B15 2TT, UK}

\author{Ismael Cognard}
\affiliation{Laboratoire de Physique et Chimie de l'Environnement et de l'Espace -- Universit\'e d'Orl\'eans / CNRS, F-45071 Orl\'eans Cedex 02, France}
\affiliation{Station de radioastronomie de Nan\c{c}ay, Observatoire de Paris, CNRS/INSU, F-18330 Nan\c{c}ay, France}

\author[0000-0002-2185-1790]{Megan E. DeCesar}
\affiliation{George Mason University, resident at the Naval Research Laboratory, Washington, DC 20375, USA}

\author[0000-0002-6664-965X]{Paul B. Demorest}
\affiliation{National Radio Astronomy Observatory, 1003 Lopezville Rd., Socorro, NM 87801, USA}

\author[0000-0001-8885-6388]{Timothy Dolch}
\affiliation{Department of Physics, Hillsdale College, 33 E. College Street, Hillsdale, MI 49242, USA}
\affiliation{Eureka Scientific, 2452 Delmer Street, Suite 100, Oakland, CA 94602-3017, USA}

\author[0000-0003-4098-5222]{F.~Adam Dong}
\affiliation{Department of Physics \& Astronomy, University of British Columbia, 6224 Agricultural Road, Vancouver, BC V6T 1Z1, Canada}

\author{Justin A. Ellis}
\altaffiliation{Infinia ML, 202 Rigsbee Avenue, Durham NC, 27701}

\author[0000-0002-2223-1235]{Robert D. Ferdman}
\affiliation{School of Chemistry, University of East Anglia, Norwich, NR4 7TJ, United Kingdom}

\author[0000-0001-7828-7708]{Elizabeth C. Ferrara}
\affiliation{Department of Astronomy, University of Maryland, College Park, MD 20742}
\affiliation{Center for Research and Exploration in Space Science and Technology, NASA/GSFC, Greenbelt, MD 20771}
\affiliation{NASA Goddard Space Flight Center, Greenbelt, MD 20771, USA}

\author[0000-0001-5645-5336]{William Fiore}
\affiliation{Department of Physics and Astronomy, West Virginia University, P.O. Box 6315, Morgantown, WV 26506, USA}
\affiliation{Center for Gravitational Waves and Cosmology, West Virginia University, Chestnut Ridge Research Building, Morgantown, WV 26505, USA}

\author[0000-0001-6166-9646]{Nate Garver-Daniels}
\affiliation{Department of Physics and Astronomy, West Virginia University, P.O. Box 6315, Morgantown, WV 26506, USA}
\affiliation{Center for Gravitational Waves and Cosmology, West Virginia University, Chestnut Ridge Research Building, Morgantown, WV 26505, USA}

\author[0000-0001-8158-683X]{Peter A. Gentile}
\affiliation{Department of Physics and Astronomy, West Virginia University, P.O. Box 6315, Morgantown, WV 26506, USA}
\affiliation{Center for Gravitational Waves and Cosmology, West Virginia University, Chestnut Ridge Research Building, Morgantown, WV 26505, USA}

\author[0000-0003-1884-348X]{Deborah C. Good}
\affiliation{Department of Physics, University of Connecticut, 196 Auditorium Road, U-3046, Storrs, CT 06269-3046, USA}
\affiliation{Center for Computational Astrophysics, Flatiron Institute, 162 5th Avenue, New York, NY 10010, USA}

\author[0000-0002-9049-8716]{Lucas Guillemot}
\affiliation{Laboratoire de Physique et Chimie de l'Environnement et de l'Espace -- Universit\'e d'Orl\'eans / CNRS, F-45071 Orl\'eans Cedex 02, France}
\affiliation{Station de radioastronomie de Nan\c{c}ay, Observatoire de Paris, CNRS/INSU, F-18330 Nan\c{c}ay, France}

\author[0000-0003-1082-2342]{Ross J. Jennings}
\altaffiliation{NANOGrav Physics Frontiers Center Postdoctoral Fellow}
\affiliation{Department of Physics and Astronomy, West Virginia University, P.O. Box 6315, Morgantown, WV 26506, USA}
\affiliation{Center for Gravitational Waves and Cosmology, West Virginia University, Chestnut Ridge Research Building, Morgantown, WV 26505, USA}

\author[0000-0001-6607-3710]{Megan L. Jones}
\affiliation{Center for Gravitation, Cosmology and Astrophysics, Department of Physics, University of Wisconsin-Milwaukee,\\ P.O. Box 413, Milwaukee, WI 53201, USA}

\author[0000-0001-6295-2881]{David L. Kaplan}
\affiliation{Center for Gravitation, Cosmology and Astrophysics, Department of Physics, University of Wisconsin-Milwaukee,\\ P.O. Box 413, Milwaukee, WI 53201, USA}

\author[0000-0001-9345-0307]{Victoria M. Kaspi}
\affiliation{Department of Physics, McGill University, 3600  University St., Montreal, QC H3A 2T8, Canada}
\affiliation{McGill Space Institute, 3550 Rue University, Montreal, Quebec H3A 2A7 Canada}

\author[0000-0002-0893-4073]{Matthew Kerr}
\affiliation{Space Science Division, Naval Research Laboratory, Washington, DC 20375-5352, USA}

\author[0000-0002-8139-8414]{Aida Yu. Kirichenko}
\affiliation{Universidad Nacional Aut\'{o}noma de M\'{e}xico, Instituto de Astronom\'{i}a, AP 106,  Ensenada 22800, BC, M\'{e}xico}
\affiliation{Ioffe Institute, Politekhnicheskaya 26, St. Petersburg, 194021, Russia}

\author[0000-0003-0721-651X]{Michael T. Lam}
\affiliation{School of Physics and Astronomy, Rochester Institute of Technology, Rochester, NY 14623, USA}
\affiliation{Laboratory for Multiwavelength Astrophysics, Rochester Institute of Technology, Rochester, NY 14623, USA}

\author[0000-0003-1301-966X]{Duncan R. Lorimer}
\affiliation{Department of Physics and Astronomy, West Virginia University, P.O. Box 6315, Morgantown, WV 26506, USA}
\affiliation{Center for Gravitational Waves and Cosmology, West Virginia University, Chestnut Ridge Research Building, Morgantown, WV 26505, USA}

\author[0000-0001-5373-5914]{Jing Luo}
\altaffiliation{Deceased}
\affiliation{Department of Astronomy \& Astrophysics, University of Toronto, 50 Saint George Street, Toronto, ON M5S 3H4, Canada}

\author[0000-0001-5229-7430]{Ryan S. Lynch}
\affiliation{Green Bank Observatory, P.O. Box 2, Green Bank, WV 24944, USA}

\author[0000-0001-5481-7559]{Alexander McEwen}
\affiliation{Center for Gravitation, Cosmology and Astrophysics, Department of Physics, University of Wisconsin-Milwaukee,\\ P.O. Box 413, Milwaukee, WI 53201, USA}

\author[0000-0002-2885-8485]{James W. McKee}
\affiliation{E.A. Milne Centre for Astrophysics, University of Hull, Cottingham Road, Kingston-upon-Hull, HU6 7RX, UK}
\affiliation{Centre of Excellence for Data Science, Artificial Intelligence and Modelling (DAIM), University of Hull, Cottingham Road, Kingston-upon-Hull, HU6 7RX, UK}

\author[0000-0002-4642-1260]{Natasha McMann}
\affiliation{Department of Physics and Astronomy, Vanderbilt University, 2301 Vanderbilt Place, Nashville, TN 37235, USA}

\author[0000-0001-8845-1225]{Bradley W. Meyers}
\affiliation{Department of Physics and Astronomy, University of British Columbia, 6224 Agricultural Road, Vancouver, BC V6T 1Z1, Canada}
\affiliation{International Centre for Radio Astronomy Research, Curtin University, Bentley, WA 6102, Australia}

\author[0000-0002-9225-9428]{Arun Naidu}
\affiliation{University of Oxford, Sub-Department of Astrophysics, Denys Wilkinson Building, Keble Road, Oxford, OX1 3RH, United Kingdom}

\author[0000-0002-3616-5160]{Cherry Ng}
\affiliation{Dunlap Institute for Astronomy and Astrophysics, University of Toronto, 50 St. George St., Toronto, ON M5S 3H4, Canada}

\author[0000-0002-6709-2566]{David J. Nice}
\affiliation{Department of Physics, Lafayette College, Easton, PA 18042, USA}

\author[0000-0002-4140-5616]{Aditya Parthasarathy}
\affiliation{Max Planck Institute for Radio Astronomy, Auf dem H \ "{u} gel 69, D-53121 Bonn, Germany}

\author[0000-0001-5465-2889]{Timothy T. Pennucci}
\affiliation{Institute of Physics and Astronomy, E\"{o}tv\"{o}s Lor\'{a}nd University, P\'{a}zm\'{a}ny P. s. 1/A, 1117 Budapest, Hungary}

\author[0000-0002-8509-5947]{Benetge B. P. Perera}
\affiliation{Arecibo Observatory, HC3 Box 53995, Arecibo, PR 00612, USA}

\author[0000-0002-8826-1285]{Nihan S. Pol}
\affiliation{Department of Physics and Astronomy, Vanderbilt University, 2301 Vanderbilt Place, Nashville, TN 37235, USA}

\author[0000-0002-2074-4360]{Henri A. Radovan}
\affiliation{Department of Physics, University of Puerto Rico, Mayag\"{u}ez, PR 00681, USA}

\author[0000-0001-5799-9714]{Scott M. Ransom}
\affiliation{National Radio Astronomy Observatory, 520 Edgemont Road, Charlottesville, VA 22903, USA}

\author[0000-0002-5297-5278]{Paul S. Ray}
\affiliation{Space Science Division, Naval Research Laboratory, Washington, DC 20375-5352, USA}

\author[0000-0002-6730-3298]{Ren\'{e}e Spiewak}
\affiliation{Jodrell Bank Centre for Astrophysics, University of Manchester, Manchester, M13 9PL, United Kingdom}

\author[0000-0002-7283-1124]{Brent J. Shapiro-Albert}
\affiliation{Department of Physics and Astronomy, West Virginia University, P.O. Box 6315, Morgantown, WV 26506, USA}
\affiliation{Center for Gravitational Waves and Cosmology, West Virginia University, Chestnut Ridge Research Building, Morgantown, WV 26505, USA}
\affiliation{Giant Army, 915A 17th Ave, Seattle WA 98122}

\author[0000-0001-9784-8670]{Ingrid H. Stairs}
\affiliation{Department of Physics and Astronomy, University of British Columbia, 6224 Agricultural Road, Vancouver, BC V6T 1Z1, Canada}

\author[0000-0002-7261-594X]{Kevin Stovall}
\affiliation{National Radio Astronomy Observatory, 1003 Lopezville Rd., Socorro, NM 87801, USA}

\author[0000-0002-1075-3837]{Joseph K. Swiggum}
\altaffiliation{NANOGrav Physics Frontiers Center Postdoctoral Fellow}
\affiliation{Department of Physics, Lafayette College, Easton, PA 18042, USA}

\author[0000-0001-7509-0117]{Chia Min Tan}
\affiliation{Department of Physics, McGill University, 3600 rue University, Montr\'eal, QC H3A 2T8, Canada}
\affiliation{McGill Space Institute, McGill University, 3550 rue University, Montr\'eal, QC H3A 2A7, Canada}

\author[0000-0003-2548-2926]{Shriharsh P. Tendulkar}
\affiliation{Department of Astronomy and Astrophysics, Tata Institute of Fundamental Research, Mumbai, 400005, India}
\affiliation{National Centre for Radio Astrophysics, Post Bag 3, Ganeshkhind, Pune, 411007, India}

\author[0000-0001-9678-0299]{Haley M. Wahl}
\affiliation{Department of Physics and Astronomy, West Virginia University, P.O. Box 6315, Morgantown, WV 26506, USA}
\affiliation{Center for Gravitational Waves and Cosmology, West Virginia University, Chestnut Ridge Research Building, Morgantown, WV 26505, USA}

\author[0000-0001-5105-4058]{WeiWei Zhu}
\affiliation{National Astronomical Observatories, Chinese Academy of Science, 20A Datun Road, Chaoyang District, Beijing 100012, China}

\begin{abstract}
    In this study, we use the Bayesian methods in the \texttt{Enterprise} \ark{package} to examine the fully general parameterization of pulsar timing models in tandem with noise.
    We investigate four pulsars\ark{, PSR J1600--3053, PSR J2043+1711, PSR J0740+6620, and PSR J1640+2224, }through the lens of Bayesian timing.
    These four are selected as they are well\ark{-}studied, but  exhibit interesting characteristics under the lens of Bayesian timing.
    We generally find good \ark{agreement between best-fit and maximum a posteriori values}, and in many cases \ark{more constrained} parameter \ark{estimations}, when comparing our method to various other Bayesian and frequentist generalized timing methods.
    We \ark{find tighter constraints on} the estimated pulsar mass for three pulsars when using the most up-to-date data releases as compared to their published results.
    Our new pulsar mass constraints (medians and 68\% confidence intervals) for our fully general \ark{non-linear} Bayesian timing models are $m_{\mathrm{p}}=1.6(1)~\mathrm{M}_{\odot}$ for PSR J2043+1711 and $m_{\mathrm{p}}=2.3^{+0.9}_{-0.7}~\mathrm{M}_{\odot}$ for PSR J1600--3053 both using the NANOGrav 12.5-yr data release, and $m_{\mathrm{p}}=2.06(6)~\mathrm{M}_{\odot}$ for PSR J0740+6620 using the data from \cite{Fonseca2021}.
    We investigate the effects on placing physical priors on timing model parameters, including restricting the upper limit on the pulsar mass for PSR J1640+2224, which has a mass often estimated to be greater than $3~\mathrm{M}_{\odot}$.
    We find \ark{that restricting the allowed sampling space of the pulsar mass for PSR J1640+2224 to} $m_{\mathrm{p}}<3~\mathrm{M}_{\odot}$ results in a pulsar mass of $m_{\mathrm{p}}=2.2(5)~\mathrm{M}_{\odot}$ for PSR J1640+2224 using the NANOGrav 12.5-yr data release.
    For the first time, we find evidence for intrinsic red noise in PSR J2043+1711.
    We show how fully general Bayesian timing can better model the interplay of the intrinsic noise and the timing parameters.
\end{abstract}

\section{Introduction}
\label{sec:Intro}

The unique physics of millisecond pulsars, and our ability to time them so precisely, makes pulsar timing a field ripe with scientific frontiers ranging from analyzing correlated signals between many pulsars in search of gravitational waves to determining the exotic makeup of neutron stars themselves \citep{Jacoby2005,TEMPO2II,Verbiest2008,15yr_gwb}.
Underlying every use of pulsar timing is the timing model constructed to account for the many different factors that can affect the pulse times of arrival (TOAs).
Pulsar timing is traditionally done by creating a model that includes a subset of all the known possible effects, and incrementally and iteratively adding or subtracting effects which have well-determined responses on pulsar TOAs.
Typically the parameters are estimated by generalized-least-squares (GLS) fitting techniques, which take into account the interplay (or correlations) between different timing parameters to minimize the \ark{chi-squared between the model and the} residuals (the model TOAs subtracted from the observed TOAs; \cite{Demorest2007, Coles2011}).
Software such as TEMPO \citep{Nice2015}, TEMPO2 \citep{TEMPO2I,TEMPO2II,TEMPO2III}, and PINT \citep{PINT, Susobhanan2024} are commonly-used tools to perform this least-squares fitting. 

Using the GLS fitting technique on many combinations of timing model parameters in tandem with other statistical tests, one can determine an optimal combination of parameters and their values. 
Generally, an F-test \ark{or the Akaike information criterion (AIC) are} used to test whether additional parameters are significant \citep{9yr_timing, Susobhanan2024}.

The algorithms currently used for gravitational wave searches in pulsar timing array (PTA) data assume that any remaining timing residuals after the initial GLS fitting can be \ark{represented by} linear corrections to the original timing solutions \citep{Jenet2009,Hobbs2009,Janssen2008,Hobbs2010,9yr_gwb}.
This assumption allows for analytical marginalization over the \emph{linearized} timing parameters when computing the likelihood used in Bayesian searches.
By marginalizing over these ``nuisance'' timing parameters, from the perspective of searching for gravitational waves, one can \ark{reduce the dimensionality of} already large parameter space, but still allow changes to the timing model to contribute to the overall likelihood.
The trade-off of this marginalization is that the linearized timing model parameters are \ark{not always accurately tracked} since any changes are assumed to be within the \ark{linear regime near} the original solutions and approximately true for any additional noise or source. See \S\ref{subsec:BayesianTiminginEnterprise} for more details.
This \ark{marginalization schema} has the potential to hide \ark{combinations of} timing parameters that may not obey the linear approximation \footnote{The \texttt{Enterprise} \citep{Enterprise, Enterprise2024} software stack can retrieve realizations of these linearized timing model parameters. We investigate differences between the recovered posteriors of the analytically-marginalized, linearized timing model parameters and the numerically marginalized parameters in Appendix~\ref{sec:appendixB}.}. 

To investigate how \ark{analytical marginalization affects} the other measurements, such as those of other sources of noise and/or gravitational waves, we implement these methods and develop sampling strategies for large data sets in \ark{the} PTA analysis package \ark{\texttt{Enterprise} and \texttt{enterprise\_extensions} \citep{Enterprise, Enterprise2024, enterpriseextensions}}. 
Our methods allow us to simultaneously account for the timing model and include noise modeling that can mitigate incorrect absorption of noise into timing model parameters.
Additionally, this generalized timing may better model timing parameters not \ark{well-approximated} in the linear regime. 
Examining the changes in the noise and timing parameter posteriors when using combinations of analytically-marginalized, \ark{linearized,} and full timing methods can elucidate which parameters obey the linear approximation and which need to be treated non-linearly.
While other works have explored this topic \citep{Lentati2014,Vigeland2014,PAL2,Susobhanan2025}, our approach explores the relationship between various noise sources and their interplay with different combinations of analytical and numerical marginalization of the timing model parameters.
In this work, we apply these techniques to individual binary millisecond pulsars timed by the NANOGrav collaboration.

\subsection{Binary Millisecond Pulsars}
\label{subsec:BinaryMillisecondPulsars}
    Pulsars with gravitationally-bound companions allow us to probe beyond the substantial amount of scientific interest provided by isolated millisecond pulsars.
    By observing binary millisecond pulsars, we can obtain information about binary formation and evolution through precise measurements of their orbital parameters, orientation, and spin properties.
    Furthermore, they are vital in tests of general relativity \citep{Barker1975,Damour1991,Damour1992}.
    
    In a non-relativistic binary system, the orbital motion can be described by Kepler's Laws alone.
    The Keplerian parameters -- the binary's orbital period, $P_{\mathrm{b}}$, \ark{the pulsar's} projected semi-major orbital axis, $x=a_{\mathrm{p}}sin~i$, orbital eccentricity, $e$, longitude of periastron, $\omega$, the epoch of periastron passage, $T_{0}$, and the position angle of the ascending node, $\Omega_{\mathrm{asc}}$ -- describe the binary system (see \cite{Lorimer2008} for a review).
    See Table \ref{tab:tm_par_def} for  timing model parameters and the \ark{variable names} we use for them.
    For relativistic binaries, post-Keplerian (PK) effects may also be measured that allow for measurements of the masses in the system and its orientation \citep{Cromartie2020}.
    
    \begin{table}[!htpb]
    \centering
    \begin{tabular}{@{} c|cc @{}}
        \hline\hline
            Label & Description & Units \\
        \hline
        $\lambda$ & Ecliptic longitude & radians \\
    	$\beta$ & Ecliptic latitude & radians \\
        RA & J2000 Right Ascension & hhmmss.ss \\
    	DEC & J2000 Declination & ddmmss.ss \\
    	$\nu$ & Spin frequency & $\mathrm{s}^{-1}$\\
    	$\dot{\nu}$ & Spin frequency derivative & $\mathrm{s}^{-2}$ \\
    	$\mu_{\lambda}$ & Proper motion in ecliptic longitude & $\mathrm{mas}~\mathrm{yr}^{-1}$ \\
    	$\mu_{\beta}$ & Proper motion in ecliptic latitude & $\mathrm{mas}~\mathrm{yr}^{-1}$ \\
        $\mu_{\mathrm{RA}}$ & Proper motion in Right Ascension & $\mathrm{mas}~\mathrm{yr}^{-1}$ \\
    	$\mu_{\mathrm{DEC}}$ & Proper motion in Declination & $\mathrm{mas}~\mathrm{yr}^{-1}$ \\
    	$\pi$ & Parallax & $\mathrm{mas}$ \\
    	$P_{\mathrm{b}}$ & Orbital period & days \\
    	$\dot{P_b}$ & Orbital period derivative & $\mathrm{days}~\mathrm{yr}^{-1}$\\
    	$T_{0}$ & Epoch of periastron & MJD \\
    	$T_\mathrm{asc}$ & Epoch of ascending node & MJD \\
    	$x$ & Projected semimajor axis & lt-sec \\
        $\dot{x}$ & Projected semimajor axis derivative & lt-sec$~\mathrm{yr}^{-1}$ \\
    	$\omega$ & Longitude of periastron & degrees \\
        $e$ & Eccentricity & -- \\
    	$\epsilon_{1}$ & $e~\sin\omega$ & -- \\
    	$\epsilon_{2}$ & $e~\cos\omega$ & -- \\
    	$i$ & Orbital inclination & -- \\
    	$m_{\mathrm{c}}$ & Companion mass & $\mathrm{M}_\odot$ \\ 
        \hline
    \end{tabular}
    \caption{Timing model parameter definitions. Each model uses a subset of all defined parameters depending on the particular model.}
    \label{tab:tm_par_def}
\end{table}
    
    The \ark{PK effect} we focus primarily on is the relativistic Shapiro delay.
    Shapiro delay occurs when the pulsar signal travels through the curved spacetime close to the companion star. This  is most easily observed at superior conjunction and for ``edge-on'' systems with inclinations $i \simeq \pi/2$.
    The measurement of Shapiro delay can lead to estimates of both the inclination of the system through the shape parameter, $s=\mathrm{sin}~i$, and the companion mass, via the range parameter, $r=\mathrm{T}_{\odot}m_{\mathrm{c}}$, where $\mathrm{T}_{\odot}=\mathrm{GM}_{\odot}/c^{3}$ \ark{as defined in \cite{Freire2010}}.
    One can examine non-General Relativistic theories of gravitation by using the measured values of the PK parameters from the Shapiro delay, Roemer delay, and \ark{Einstein delay} and compare the values predicted by the appropriate theory \citep{FonsecaThesis}. 
    The independent measurement of the companion mass can give vital insight into the system's evolutionary history, providing a test of various binary-evolution theories \citep{Fonseca2016}.
    
    In this work, we use the Shapiro delay to estimate the pulsar mass by solving the Keplerian mass function
    \begin{equation}
        \label{eq:mass_function}
        f_{m} = \frac{4\pi^{2}}{\mathrm{T}_{\odot}}\frac{x^{3}}{P_{\mathrm{b}}^{2}} = \frac{\left(m_{\mathrm{c}}\sin~i\right)^{3}}{\left(m_{\mathrm{c}}+m_{\mathrm{p}}\right)^{2}} ,
    \end{equation}
    where $x$ is the projected semimajor axis, $m_{\mathrm{p}}$ is the pulsar mass, $m_{\mathrm{c}}$ is the companion mass, and $P_{\mathrm{b}}$ is the binary period \citep{Freire2010}.
    Without an independent, non-Keplerian measurement of the binary companion mass, the mass of the pulsar cannot be uniquely determined.

\subsection{Neutron Star Mass}
    One of the difficulties in determining the exact composition of neutron stars stems from the wide range of maximum \ark{masses predicted by} different \ark{proposed equations of state}.
    While many advances in nuclear models have been developed in the intervening decades, there is still a large range of viable theories for the actual compositions of neutron star cores \citep{Alsing2018,Rezzolla2018}.
    These theories, and their predicted \ark{equations of state}, are all constrained by the mass-radius relation calculated by solving the general relativistic structure equations. 
    While pulsar timing cannot directly determine the radii of neutron stars, it can place tight constraints on their masses, narrowing down the region of viable theories with each new ``most-massive'' neutron star. 
    These updated masses also help probe binary evolution to understand late-stage stellar evolution and mass transfer.
    For a review of theories and constraints on the neutron star equation of state, see \cite{Ozel2016}.

\subsection{Eccentricity}
    To model the numerous binary parameters, it is common to classify binary systems through their eccentricity.
    Binary systems exhibiting \ark{eccentricity greater than $0.001$}, among other factors like the projected semimajor axis $x$, are examined using the model developed by \cite{Damour1985,Damour1986,Damour1992}, which contains all of the parameters for a relativistic binary system, hereafter known as the ``DD'' binary model.
    
    Nearly circular systems ($e<0.001$) have highly covariant parameters, specifically the epoch of periastron passage ($T_{0}$) and longitude of periastron ($\omega$).
    To account for the numerical instability caused by the covariance between the two parameters, \cite{Lange2001} developed the ``ELL1'' binary model.
    The ELL1 model, originally developed for use on PSR J1012+5307, uses an expansion of the orbital parameters in the small eccentricity approximation of the Roemer and Shapiro timing delays.
    Thus the model uses two Laplace-Lagrange parameters, $\eta = e~\mathrm{sin}\omega$ and $\kappa = e~\mathrm{cos}\omega$, and the epoch of ascending-node passage, $T_{\mathrm{asc}}$, in addition to the orbital period, $P_{\mathrm{b}}$, and the semi-major axis of the orbit projected onto the line of sight, $x$.
    Both models for eccentricity allow for the fitting of PK parameters in addition to their respective model parameters.
    For in-depth discussion on pulsars in binary systems, see \cite{Lorimer2012}.

\subsection{Dispersion Measure Modeling}
\label{subsec:IntroDMModeling}
    The interstellar medium (ISM) causes a frequency-dependent delay in radio pulses due to dispersion in the ionized ISM.
    The dispersion measure (DM) is the column density of electrons along the line of sight to the pulsar. 
    The frequency-dependent delay is linearly dependent on the DM and decreases as the inverse radio frequency squared.
    It is determined by fitting arrival times across wide bandwidths, \ark{and when using narrowband receivers, this} often necessitates using more than one radio receiver \citep{Demorest2013}.
    Over time, the electron density, and thus the measured DM, varies because of line of sight changes and inhomogeneities in the \ark{ISM and} solar wind \citep{Jones2017}.
    The variations take many forms of deterministic and stochastic behaviors \citep{Lam2016}.
    Improperly accounting for DM variations can possibly leave unmodeled low-frequency noise in the PTA datasets \citep{Lam2015,Shannon2017,Goncharov2021}.
    
    The standard method to account for variations in the DM is by including a piecewise-constant DM function dependent on time.
    This model, known as ``DMX'', is applied on roughly an epoch-by-epoch basis typically determined by the observation cadence \citep{Demorest2013}.
    For NANOGrav timing analyses, the bin size used for DMX fitting includes all the TOAs in a one-to-six day window and the DM is assumed constant within each bin.
    DMX is a powerful tool for mitigating the effects of the ever changing ISM on pulse arrival times, but since each bin is fit independently, the model can also absorb higher order scattering variations with a typically steeper frequency dependence \citep{Shapiro2021}.
    \ark{More recent studies have shown DM can often be better modeled as Fourier-domain Gaussian processes \citep{Hazboun2022,Iraci2024, Hazboun2025}.}

\subsection{Selection of Binary Millisecond Pulsars}
    The four pulsars studied in this work are selected for their long datasets and unique characteristics.
    Each pulsar exemplifies facets of commonality that can be generalized across millisecond pulsars, but also demonstrate the individual constraints and difficulties for each pulsar.
    All four pulsars are included in the NANOGrav gravitational wave search which models all pulsars as containing combinations of white and red noise (see section \S\ref{sec:Methods} for a detailed explanation) \citep{12p5yr_gwb, 12p5yr_timing}.
    PSR J2043+6620 exhibits changes in its binary parameters with the inclusion (or exclusion) of intrinsic red noise \citep{9yr_timing, 12p5yr_timing}.
    PSR J1600--3053 has an anomalously large inferred pulsar mass with poor constraints \citep{12p5yr_timing}.
    PSR J0740+6620 has a well-constrained pulsar mass near the predicted maximum mass range for neutron stars with a dataset tailored to hone in on the Shapiro delay \citep{12p5yr_timing,Cromartie2020, Fonseca2021}.
    PSR J1640+2224 has been studied with Bayesian timing in older datasets and represents a pulsar with less well-behaved timing model parameters as they have shifted by significant amounts between published datasets \citep{Demorest2013, 9yr_timing, 12p5yr_timing}.
    By examining these four pulsars we explore the difficulties and the impact on the timing model of modeling individual pulsars, and emphasize the consequences of treating all pulsars the same in a larger set.
 
\subsection{Paper Outline}      
    We apply our Bayesian timing methods to a sample of pulsars.
    In section \S\ref{sec:Methods} we discuss our methods, and in the following four sections we apply our analyses to binary pulsars PSRs J2043+1711 (\S\ref{sec:J2043}), J1600--3053 (\S\ref{sec:J1600}), J0740+6620 (\S\ref{sec:J0740}), and J1640+2224 (\S\ref{sec:J1640}). 
    In particular, we examine changes in the noise and timing parameter posteriors when using combinations of linearized and fully general timing models.
    Section \S\ref{sec:Discussion} discusses the results and implications of our new methods when applied to these binaries and other pulsar systems.
    In addition, section \S\ref{sec:Discussion} presents plans for future work and our conclusions.

\section{Methods}
\label{sec:Methods}

\subsection{Bayesian Timing in \texttt{Enterprise}}
\label{subsec:BayesianTiminginEnterprise}
    In this work, we use the Bayesian methods newly implemented in the pulsar timing analysis code \texttt{Enterprise}.
    The noise characterization and fundamental framework for \texttt{Enterprise} are well documented in \cite{9yr_gwb,11yr_gwb,12p5yr_gwb} and are used within this work to examine the noise properties of the pulsars studied.
    Each type of analysis and their combinations include noise for each pulsar.
    The white noise parameters \citep[EFAC, EQUAD, ECORR][]{5yr_bwm, 9yr_gwb} are always used and a set of the three parameters are fit for each  receiver backend.
    
    While none of the selected pulsars have red noise (RN) included in their published timing models, we use the RN parameters (amplitude, $A$, and spectral index, $\gamma$) to examine its effects on the white noise and timing model parameters.
    We closely evaluate the inclusion of RN because in typical GW detection analyses, every pulsar is given intrinsic RN parameters.
    Thus by including RN \ark{in our analyses} we can more closely examine the effects of RN on timing parameters, particularly when \ark{there is little evidence found in other works to indicate the pulsar has RN \citep{Lam2017,Lam2025}.}
    
    The main divergence of this work and previous descriptions of the capabilities of \texttt{Enterprise} lie within the timing model.
    In typical PTA analyses, the timing parameters are estimated by a generalized-least-squares (GLS) method that maximizes the likelihood of the timing model.
    Any remaining timing residuals are then assumed to be characterized by linear corrections to the original timing solutions.
    In practice, the likelihood includes a timing-model design matrix made up of derivatives of the timing model with respect to the parameters, i.e. encoding a linearized model, around the best-fit parameters.
    The design matrix consists of the partial derivatives of the timing model with respect to the timing model parameters. A first-order Taylor-series approximation of the timing model can be obtained by matrix multiplication of the design matrix with a set of basis coefficients that represent the deviations of the timing model parameters around the pre-fit parameters.
    These coefficients are given a diagonal multivariate Gaussian prior with infinite variance and no covariances. \ark{Using the methods of \cite{Lentati2013,vanHaasteren2015,9yr_gwb}, we are then able to \emph{analytically} marginalize over the timing parameters.}
    We use the analytically-marginalized, linearized method from this last reference in our models A, as shown in table \ref{tab:models}.

\subsection{Full Timing Models}
\label{subsec:NonlinearTimingModels}
    In this work, we adapt this method by explicitly including the timing model as a deterministic signal.
    The signal for the range of timing model parameters is generated by recomputing the pulsar residuals at each MCMC step and subtracting the new step from the initial residuals from the GLS fit.
    This allows us to retrieve the timing model \ark{residuals} for each move in the $N_{\mathrm{par}}$ space.
    We use both \texttt{libstempo}, a Python wrapper around TEMPO2 that is closely integrated into \texttt{Enterprise}, and \texttt{PINT}, a fully Python-based timing package, to recalculate the timing residuals using the full model, which \ark{includes both} nonlinear (e.g., binary parameters) and natively linear components (e.g., JUMPS).
    The change in delay is then incorporated into the \texttt{Enterprise} likelihood as another modification to the TOAs along with the noise and other deterministic signals.
    Model B uses these methods, as detailed in table \ref{tab:models}.

\subsubsection{Combining Full and Linearized Timing}
\label{subsec:NonlinearAndLinearTimingModels}
    Because of the necessity of determining the new residuals at each step, and the inclusion of an extra phase offset parameter necessary to align the initial phase by \texttt{libstempo}, we also allow for combinations of the analytically-marginalized, linearized model and direct sampling of the full timing model.
    To execute this, we filter the sampled timing parameters from the pulsar's design matrix, and then perform the same process as in the fully generalized analysis, where they contribute to the likelihood through the delay from changes in the residuals from their initial values to the newly sampled values.
    The non-sampled parameters are then modeled as a linear approximation (where not natively linear) and analytically marginalized by the same means as in section \S\ref{subsec:BayesianTiminginEnterprise} with unconstrained basis coefficients.
    We include this option because pulsars often have over a hundred DMX parameters,  and directly sampling these parameters can increase the computation time by hundreds of hours in some cases. 
    The DMX model is a linear model from its inception and so there is no loss of generality by using the linearized algorithm.
    We show comparisons between using fully varied timing models and the combination of marginalized ``nuisance'' parameters (DMX, frequency dependent (FD) parameters to account for pulse profile evolution, and JUMP parameters for changes in telescope backend) for all four pulsars in this study. Examples of the posteriors for these timing parameters are shown in Appendix~\ref{sec:appendixB}.
    The comparison shows that in most cases there is no change in noise and other timing parameters when Using the linearized model for DMX.
    This is evidence that these parameters can be modeled with the analytically-marginalized, linearized framework in most, if not all, cases.
    We refer to this as model C as detailed in table \ref{tab:models}.

    In theory, since both model C and model B analyses are exploring the same parameter space, the maximum a \ark{posteriori} should be identical. In practice however, the parameter space is of very high dimensionality and local minima abound, especially when \ark{inferring} binary parameters \citep{Fonseca2021}. Thus we explore the effects of marginalizing over the timing model parameters both analytically, numerically, and  in combination.

    The legacy code \texttt{PAL2} \citep{PAL2}, a precursor to \texttt{Enterprise}, uses a similar method as our model C analysis to directly sample timing model parameters, while analytically marginalizing over the many ``nuisance'' parameters of the timing model.
    \texttt{PAL2} is used in \cite{Fonseca2016} to compare to their chi-squared gridding techniques, for which there was good agreement.
    In order to complete a similar comparison to our analyses, we use \texttt{PAL2} in some of the pulsar analyses in relevant later sections.
    Because of the similarities to \texttt{Enterprise}, we are able to compare not just the timing model posteriors, but the noise parameters as well.
    
\begin{table}[!htbp]
    \centering
    \hspace{-20mm}
    \scalebox{0.9}{
    \makebox[\columnwidth]{
        \begin{tabular}{@{} c|cc|cc @{}}
            \hline\hline
               & \multicolumn{2}{c}{Timing Model} & \multicolumn{2}{c}{DMX} \\
            \hline\hline
            Model & Full & Linearized & Full & Linearized \\
            \hline
            Model A & -- & \checkmark & -- & \checkmark \\
            Model B & \checkmark & -- & \checkmark & -- \\
            Model C & \checkmark & \checkmark & -- & \checkmark \\
            \hline
        \end{tabular}
    }}
    \caption{Model descriptions used in our analyses. Each use a combination of the numerically--marginalized, fully general (Full) timing model parameters  and the analytically--marginalized, linearized (Linearized) parameters. The analytically--marginalized parameters are all DMX, all FD parameters, and all JUMP parameters for all four pulsars in the study.}
    \label{tab:models}
\end{table}

\subsection{\ark{Prior Distributions and Sampling}}
    \label{subsec:global_model_params}
    Each Bayesian analysis makes an assumption about the shape and constraints of the prior probability distribution on each sampled parameter.
    The most agnostic type of prior is a uniform, ``ignorance'' prior that does not use a narrower distribution to bias the sampled space.
    All of our priors are uniform and uninformative except in the specific case where we attempt to duplicate the priors of \cite{Vigeland2014} in the analysis of the NANOGrav 5-yr data set for J1640+2224 in section \S\ref{sec:J1640}.
    For the noise parameters, we use the standard set of red and white noise priors used in analyses with \texttt{Enterprise} found in the \texttt{enterprise\_extensions} module \cite{enterpriseextensions}.
    The timing model parameters prove to be more difficult to sample due to the complicated correlations and high level of precision of many timing parameters.
    In order to solve this we choose a reasonable range in which the timing model may vary, without overly constraining the parameters.
    We use an effective sampled space of $\left[\mu_{\mathrm{GLS}}-50\sigma_{\mathrm{GLS}},\mu_{\mathrm{GLS}}+50\sigma_{\mathrm{GLS}}\right]$ without losing precision and accuracy due to floating point and round-off errors. 
    While these may seem wide \ark{as compared to previous studies' use of $\pm10\sigma_{\mathrm{GLS}}$ \citep{Lentati2014,Susobhanan2025}}, they are still narrow compared to the effectively infinite priors of the linearized model.
    \ark{Further study may be necessary to determine the optimal width and shape of the priors for each timing model parameter.}
    \ark{Priors used in this work can be found on Table \ref{tab:priors}}

    \begin{table*}[!htb]
    \centering
    \begin{tabular}{@{}c|c|c|c|c@{}}
    Parameter               & Pulsars                & Dataset       & Prior Shape & Prior Range \\
    \hline\hline
    Timing parameters       &  All                   & All           & Uniform     & $\left[\mu_{\mathrm{GLS}}-50\sigma_{\mathrm{GLS}},~\mu_{\mathrm{GLS}}+50\sigma_{\mathrm{GLS}}\right]$ \\
    DMX, FD, and JUMP parameters &  All                   & All           & Uniform     & $\left[\mu_{\mathrm{GLS}}-50\sigma_{\mathrm{GLS}},~\mu_{\mathrm{GLS}}+50\sigma_{\mathrm{GLS}}\right]$ \\
    $\cos i$                &  All                   & All           & Uniform     & $\left[\mathrm{Max}\left(0,~\mu_{\mathrm{GLS}}-50\sigma_{\mathrm{GLS}}\right),~\mathrm{Min}\left(\mu_{\mathrm{GLS}}+50\sigma_{\mathrm{GLS}},~1\right)\right]$ \\
    $\pi$                   &  All                   & All           & Uniform     & $\left[\mathrm{Max}\left(0,~\mu_{\mathrm{GLS}}-50\sigma_{\mathrm{GLS}}\right),~\mu_{\mathrm{GLS}}+50\sigma_{\mathrm{GLS}}\right]$ \\
                            &  J1640+2224            & 5-yr          & Normal      & Eq.~\ref{eq:DM_prior} \\
    $m_{\mathrm{c}}$        &  All except J1640+2224 & All           & Uniform     & $\left[10^{-10},~\mu_{\mathrm{GLS}}+50\sigma_{\mathrm{GLS}}\right]$ \\
                            &  J1640+2224            & 5-yr          & Uniform     & $\left[10^{-10}~\mathrm{M}_{\odot},~10~\mathrm{M}_{\odot}\right]$ \\
                            &  J1640+2224            & 9-yr, 12.5-yr & Uniform     & $\left[10^{-10}~\mathrm{M}_{\odot},~1.4~\mathrm{M}_{\odot}\right]$ \\
    $e$                     &  All                   & All           & Uniform     & $\left[0,~0.9999\right]$\\
    EFAC                    &  All                   & All           & Uniform     & $\left[0.01,~10\right]$ \\
    ECORR                   &  All                   & All           & log-Uniform & $\left[-8.5,~-5\right]$ \\
    EQUAD                   &  All                   & All           & log-Uniform & $\left[-8.5,~-5\right]$ \\
    $A_{\mathrm{RN}}$       &  All                   & All           & log-Uniform & $\left[-20,~-11\right]$ \\
    $\gamma_{\mathrm{RN}}$  &  All                   & All           & Uniform     & $\left[0,~7\right]$ \\
    \end{tabular}
    \caption{Prior definitions used in each model. "All`` for the Pulsars column refer to PSR J1600--3053, PSR J2043+1711, PSR J0740+6620, and PSR J1640+2224, while "All`` for the Dataset column refers to the NANOGrav 12.5-yr dataset for all pulsars \citep{12p5yr_timing}, the NANOGrav 5-yr dataset for PSR J1640+2224 \citep{Demorest2013}, the NANOGrav 9-yr dataset for PSR J2043+1711 and PSR J1640+2224 \citep{9yr_timing}, and to the data from \cite{Cromartie2020} and \cite{Fonseca2021} for PSR J0740+6620. Timing parameters not explicitly mentioned in the table all are combined in the first row. DMX, FD, and JUMP parameters when sampled in Model B analyses are combined in the second row.}
    \label{tab:priors}
\end{table*}

    Some timing model parameters span values that are unphysical.
    To ensure that parameters remain in physical ranges, we set hard boundaries on the priors of eccentricity, inclination, parallax lower bound, and companion mass lower bound for all pulsars.
    \ark{For inclination, we uniformly sample in $\cos i$ because a collection of randomly-oriented binary systems possesses a uniform distribution in cosine of the inclination.}
    Furthermore, for pulsar J1640+2224, we impose a more physical range upon the derived pulsar mass parameter given in equation \ref{eq:mass_function} of a maximum mass of $3~\mathrm{M}_{\odot}$.
    While current equations of state place an upper limit on the mass of neutron stars of   $m_{\mathrm{p}}<2.6~\mathrm{M}_{\odot}$ \citep{Alsing2018,Rezzolla2018}, we choose a more conservative maximum mass cutoff of $3~\mathrm{M}_{\odot}$ due to the uncertainties.
    \ark{We simultaneously restrict the prior on the companion mass to be uniform between $10^{-10}$ and the Chandrasekhar limit of $1.4~\mathrm{M}_{\odot}$ as the companion mass has been observed to be a WD (see Section \S\ref{sec:J1640}) \citep{Lundgren1996,Lohmer2005}.}
    We examine the effect of the maximum pulsar mass constraint in section \S\ref{sec:J1640}.
    
    In the analysis of the NANOGrav 5-yr data set for J1640+2224, we also use an informative prior on the parallax distance. The GLS best-fit value is consistent with zero and thus is not significant in the fit, but parallax affects all pulsars and thus was included in the final GLS fit.
    To directly compare to the work of \cite{Vigeland2014}, we use the same prior on the parallax
    \begin{equation}
        \label{eq:DM_prior}
        p(\mathrm{PX})=\frac{1}{\sqrt{2 \pi} \sigma_{d} \mathrm{PX}^2} \exp \left[-\frac{\left(\mathrm{PX}^{-1}-d\right)^2}{2 \sigma_d^2}\right], \quad \mathrm{PX}>0 ,
    \end{equation}
    based upon the distance estimation from the pulsar's DM from \cite{Cordes2002}, where $d$ is the assumed DM distance, and $\sigma_{d}^{2}$ is the variance of the Gaussian distribution for $d$, which is set to $\sigma_{d}=0.2d$, as in \cite{Vigeland2014}.
    
    For all of the analyses on each model, we found that initializing the timing model parameters in our sampler (PTMCMCSampler from \cite{PTMCMC}) with the best-fit values from the GLS analyses used to produce the parameter files for each data set allowed the sampler to converge more quickly.
    While it is possible to randomly start in the given prior for all parameters, because of the large exploration space and the covariance between timing model parameters, it becomes much slower for the posteriors to converge.

    \ark{Furthermore, to improve sampling speed and efficiency, we use PTMCMCSampler's implementation of parallel tempering in all runs as well as ``jump proposals", proposal distributions from which to sample, described in section III C. of \cite{Enterprise2024}.}
    \ark{We construct additional jump proposals specifically for sampling in the timing parameters by drawing samples from their priors in groups of parameters that would all change together (i.e. position, spin, Keplerian parameters, etc.).}
    
    We report all results for chains that have auto-correlation lengths $<$ 200, and a Gelman-Rubin split R hat statistic $<$ 1.1 \citep{Vehtari2019} except in specifically mentioned cases where convergence requires more computational resources than feasible, see section \S\ref{sec:J1640}.

\section{PSR J2043+1711}
\label{sec:J2043}

Pulsar J2043+1711 was discovered by the \textit{Fermi} gamma-ray telescope with its Large Area Telescope (LAT) instrument as an unassociated gamma-ray source \citep{Atwood2009}.
It was discovered at  radio wavelengths by \cite{Guillemot2012} as a 2.4-ms pulsar in a 1.5-day orbit with a low-mass companion using the Nan\c{c}ay telescope and with confirmation by the Green Bank Telescope.
\cite{Guillemot2012} then constructed a timing model with the Nan\c{c}ay, Westerbork, and Arecibo telescopes, and were able to detect the Shapiro delay.
While their Shapiro delay measurements was not significant enough to yield statistically meaningful estimates of the pulsar and companion masses or the system's inclination angle, they \ark{placed} constraints on the parameters by \ark{assuming that} the relation between $m_{\mathrm{p}}$ and $P_{\mathrm{b}}$ derived by \cite{Tauris1999} applies.
The assumption of the system being a close millisecond pulsar/He-WD binary allowed them to set limits of $0.20 < m_{\mathrm{c}} < 0.22~\mathrm{M}_{\odot}$ on the companion mass, $1.7 < m_{\mathrm{p}} < 2.0~\mathrm{M}_{\odot}$ on the pulsar mass, and $i = 81.3(1.0)$ \ark{degrees} on the system inclination.
Using the targeted Shapiro-delay observations in the NANOGrav 9-yr data set, \cite{9yr_timing} found \ark{improved measurements} of the Shapiro delay.

The findings in the 9-yr data set allowed \cite{Pennucci2015} to place improved measurements of $m_{\mathrm{c}} = 0.170^{+0.012}_{\text{--}0.012}~\mathrm{M}_{\odot}$, and $i = 83.5^{+1.4}_{\text{--}1.7}$ degrees using probability maps generated from $\chi^{2}$-gridding techniques from \cite{Splaver2002}.
Using the maximum likelihood points for these two values, they estimate the pulsar mass to be $m_{\mathrm{p}} = 1.35~\mathrm{M}_{\odot}$, which is in agreement with the timing model presented in \cite{Guillemot2012}, but not with the assumption of the $P_{\mathrm{b}}$--$m_{\mathrm{p}}$ relation ($m_{\mathrm{p}}>1.7~\mathrm{M}_{\odot}$). 

Using the same NANOGrav 9-yr data, \cite{Fonseca2016} compared the $\chi^{2}$-gridding with a fully Bayesian method, including white noise, using \texttt{PAL2} and found the companion mass to be $m_{\mathrm{c}} = 0.175^{+0.016}_{\text{--}0.015}~\mathrm{M}_{\odot}$, and the inclination angle to be $i = 83.2^{+0.8}_{\text{--}0.9}$ \ark{degrees}; they estimate the pulsar mass to be $m_{\mathrm{p}} = 1.41^{+0.21}_{\text{--}0.18}~\mathrm{M}_{\odot}$.
Both the \cite{Pennucci2015} and \cite{Fonseca2016} results showed good agreement for the Shapiro delay parameters.

In this work, we use the NANOGrav 9-yr data to compare previous Bayesian results from \texttt{PAL2} and $\chi^{2}$-gridding analyses in \cite{Fonseca2016}. 
Then we use the NANOGrav 12.5-yr data to perform the first fully general Bayesian analysis on J2043+1711. 

\begin{figure*}
    \centering
    \includegraphics[width=\textwidth]{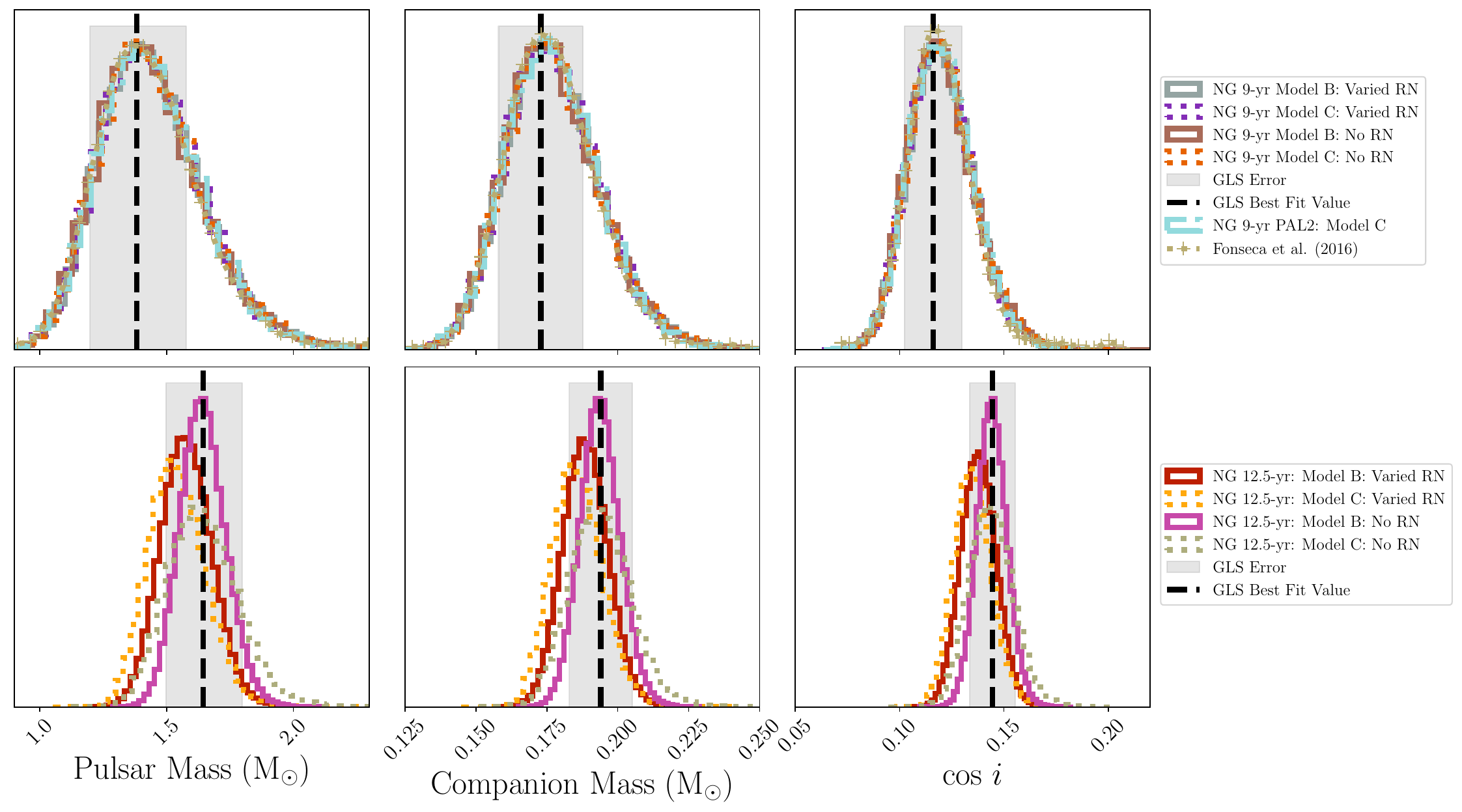}
    \caption{Shapiro delay parameter plots for PSR J2043+1711. 
    The top three panels contain the results of a model B (solid blue), a model C (dashed orange), a model C \texttt{PAL2} analysis (solid red), and the results from the chi-squared gridding analysis of \cite{Fonseca2016} (pink dotted crosses) on the NANOGrav 9-yr dataset. 
    The bottom three panels contain the results for the NANOGrav 12.5-yr dataset using a model B (solid tan), and a model C (dashed light blue) analysis. 
    The dashed black line represents the best fit value from the GLS analysis of each dataset (the top being \cite{9yr_timing} and the bottom \cite{12p5yr_timing}) with the shaded region being their respective one-sigma error regions. 
    None of the represented analyses include RN.
    }
    \label{fig:J2043_Mass_Plots}
\end{figure*}

As seen in figure \ref{fig:J2043_Mass_Plots}, the posteriors for both model B and C are nearly identical to the results from \cite{Fonseca2016} and a \texttt{PAL2} model C analysis done for comparison. 
This similarity also holds between the results from the GLS analysis in \cite{9yr_timing}.
Both the model B and C analyses use no RN so as to closely replicate the analyses done in \cite{Fonseca2016} and \cite{9yr_timing}.
We find that there is virtually no statistical difference in the timing model posterior distributions between our methods and those of previous studies using the same data.

\subsection{Noise Analysis}
    We perform the same analysis on the NANOGrav 12.5-yr data set, and as there have been no Bayesian analyses yet done on the timing model parameters, we compare our results to those from the GLS analyses in \cite{12p5yr_timing}.
    In addition to the comparison between model B and C with no RN, we perform the same two analyses with intrinsic RN included despite there being insufficient evidence for the inclusion of RN in J2043+1711 in the analyses of \cite{12p5yr_timing}.
    
    We find that for the analyses without RN, model B and model C timing model parameter posteriors are consistent with each other.
    All median posterior values (shown in table \ref{tab:J2043_tm_changes_pt1} and \ref{tab:J2043_tm_changes_pt2}) are within the one-sigma error regions of the GLS analysis.

    \begin{figure*}[!htbp]
        \centering
        \includegraphics[width=\textwidth]{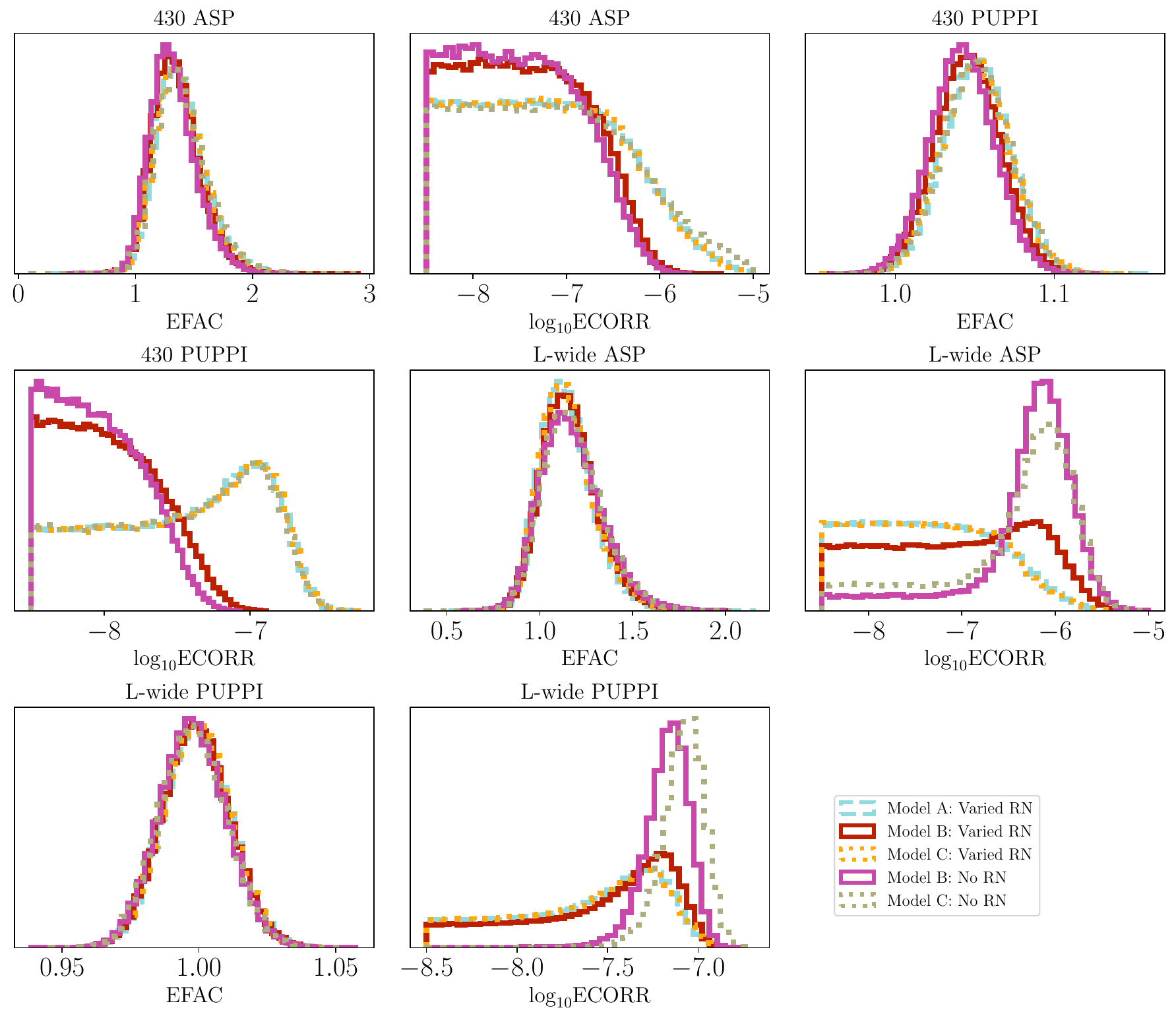}
        \caption{White noise parameter posteriors separated by observing backends for PSR J2043+1711 using the NANOGrav 12.5-yr dataset \citep{12p5yr_timing}.
        The histograms show model A with RN as the dashed cyan lines, model B with and without RN as the solid red and solid magenta lines, respectively, and model C with and without RN as the dotted orange and dotted light olive--green lines, respectively.
        }
        \label{fig:J2043_Noise_Plots}
    \end{figure*}

    Model C posteriors are broader distributions than found in \cite{12p5yr_timing}, while model B results in improved constraints in many of the timing model parameters.
    There is a more evident difference between model B and C when examining the noise parameters as shown in figure \ref{fig:J2043_Noise_Plots}.
    Most noticeably, the ECORR value is slightly more significant when using the linearized model for the parameters, whereas in the fully general case, the ECORR shows preference for values $\mathrm{log}_{10}(\mathrm{ECORR}) < -8.5$.
    
\begin{figure*}
        \centering
        \includegraphics[width=\textwidth]{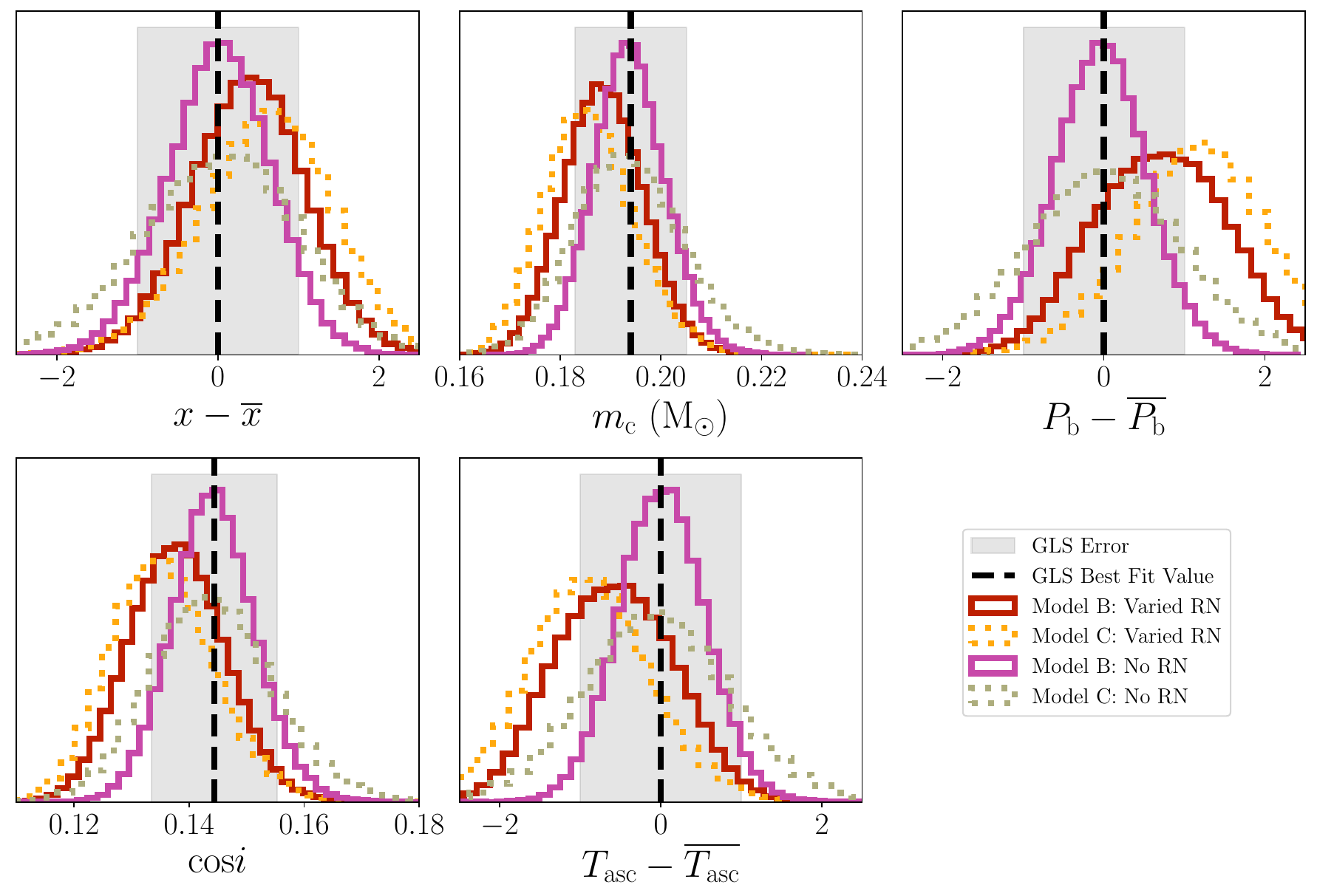}
        \caption{Comparison of Keplerian parameters of PSR J2043+1711 for model B with and without RN, shown by the solid red and solid magenta histograms, respectively and model C with and without RN, shown by the dotted orange and dotted light olive--green histograms, respectively.
        Each analysis is done on the NANOGrav 12.5-yr dataset \citep{12p5yr_timing}.
        The dashed black line represents the best fit value from the GLS analysis of \cite{12p5yr_timing} with the shaded region being the respective one-sigma error regions.
        }
        \label{fig:J2043_kep_comp_plots}
    \end{figure*}

    When we include intrinsic RN in model B and C, we find that the timing model parameters are in most cases within the one-sigma error region of the GLS analyses, but the posterior is no longer centered on the best-fit value from the original timing analyses.
    These shifts are evident from figure \ref{fig:J2043_kep_comp_plots}, for a sampling of the Keplerian parameters for J2043+1711.
    
\begin{figure}
        \centering
        \includegraphics[width=\columnwidth]{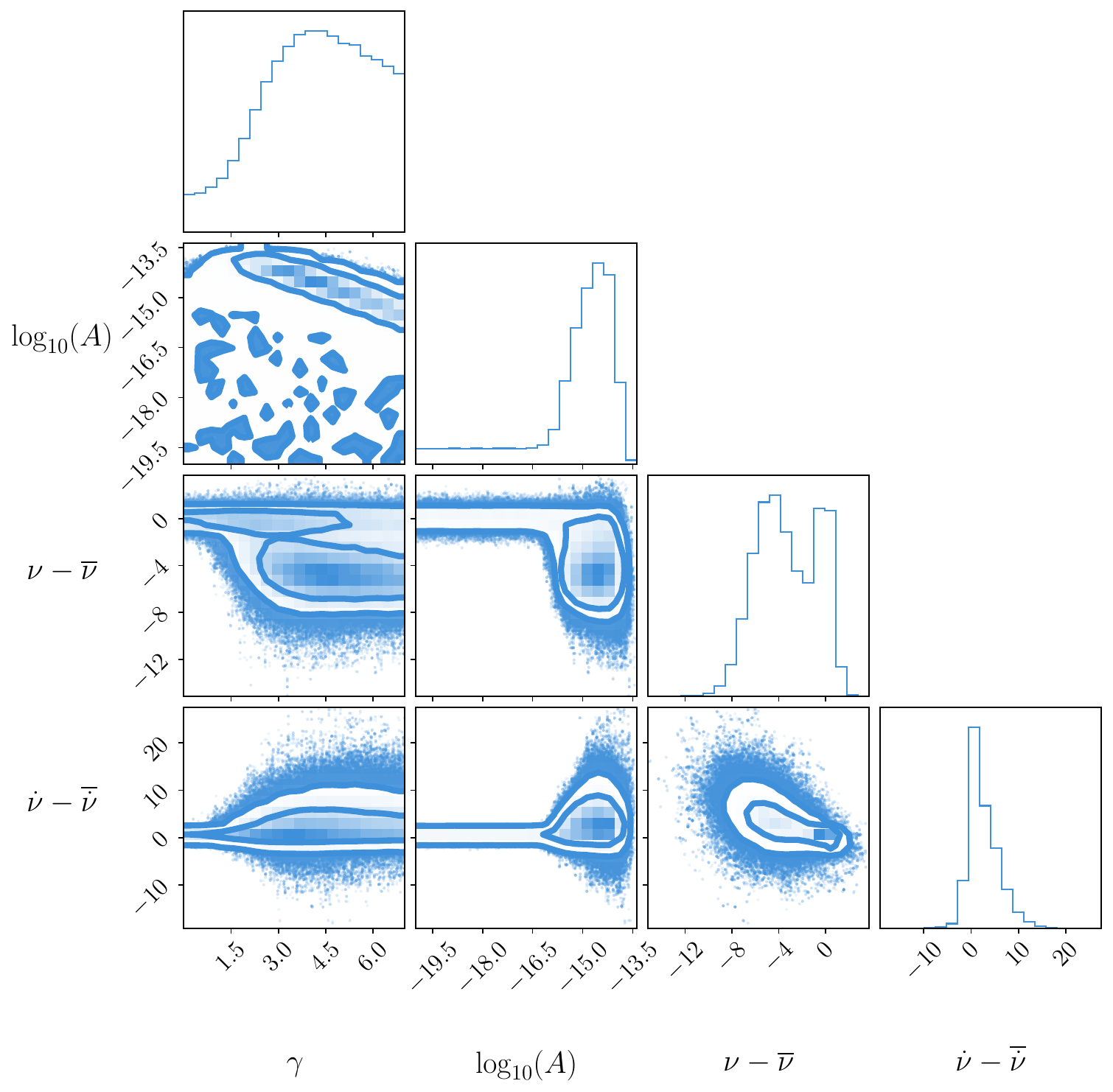}
        \caption{Corner plot of RN versus spin parameters for PSR J2043+1711 in the NANOGrav 12.5-yr dataset \citep{12p5yr_timing} for the model B analysis.
        The histogram panels show the 1-D marginalized posteriors and the contour plots show the 2-D marginalized posteriors for the relevant parameters. Note that $\nu$ and $\dot{\nu}$ are shown in number of error bars away from the mean, as determined from the GLS fit.
        }
        \label{fig:J2043_RN_spin_corner_plots}
    \end{figure}

    The furthest shifts from the results of \cite{12p5yr_timing} when including RN occur in the rotational frequency and its first time derivative.
    In figure \ref{fig:J2043_RN_spin_corner_plots}, we show the direct impact of the RN parameters on $\nu$ and $\dot{\nu}$.
    As the RN increases in amplitude, both $\nu$ and $\dot{\nu}$ become more uncertain; see table \ref{tab:J2043_tm_changes_pt1} and \ref{tab:J2043_tm_changes_pt2} for detailed values for each.
    We also note the now significant RN present in the pulsar, with a Savage-Dickey Bayes' factor (SDBF, \cite{Dickey1971}) of $\mathrm{SDBF}=2.6$ for model B and $\mathrm{SDBF}=180$ for model C, and a clearly red spectral index preferring a steeper RN process.
    
    Because of the presence on a common spectral process in the NANOGrav 12.5-yr data set, we also attempted an analysis with an additional RN process with a freely varying amplitude corresponding to a fixed spectral index on $\gamma=13/3$, the spectral index predicted for a stochastic gravitational wave background made up of \ark{circular supermassive} black hole binaries \citep{12p5yr_gwb}.
    
    \begin{figure}
        \centering
        \includegraphics[width=\columnwidth]{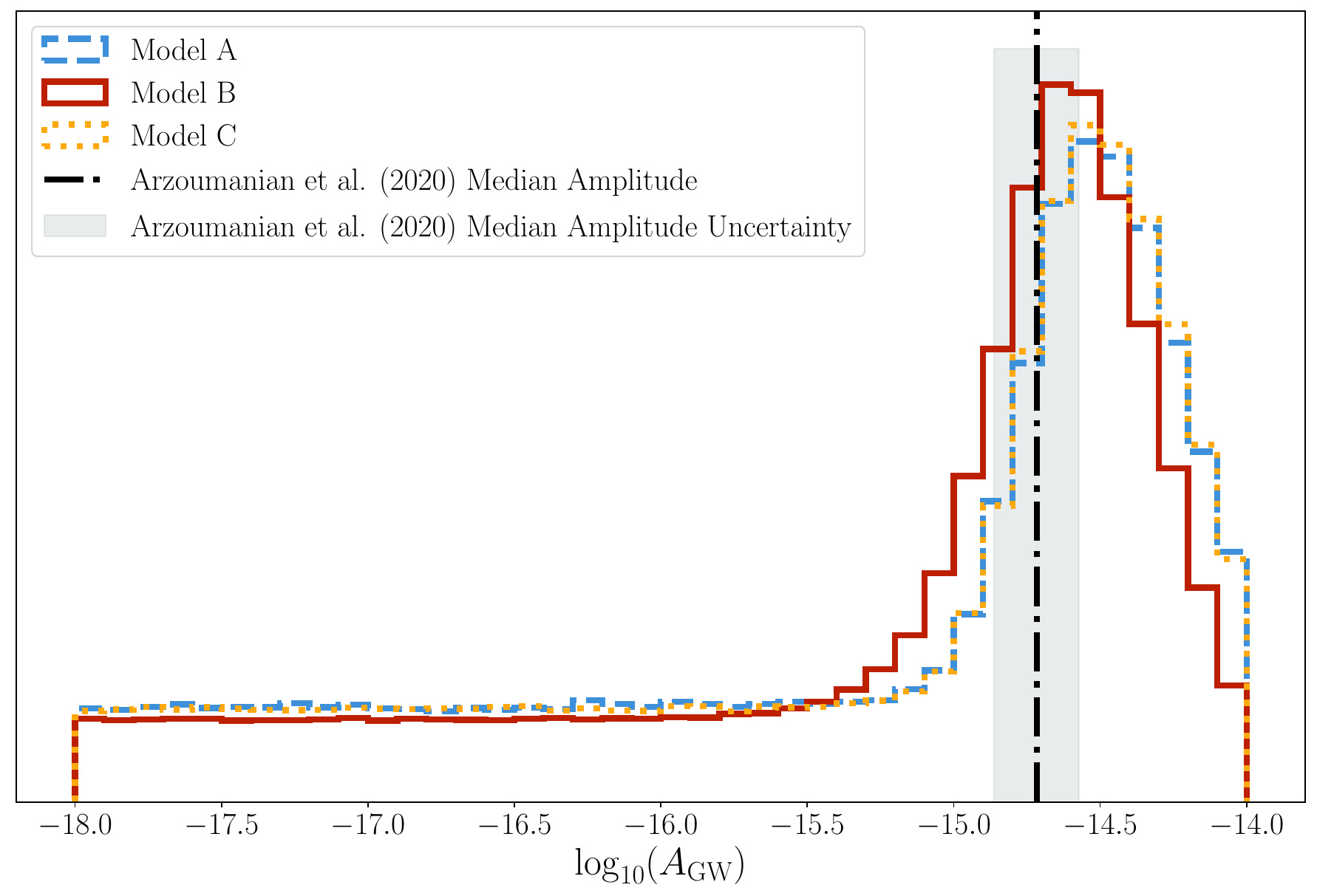}
        \caption{Amplitude of a $\gamma=13/3$ process in addition to a model A (dashed blue), B (solid red), and C (dashed orange) analysis with intrinsic RN included for PSR J2043+1711 in the NANOGrav 12.5-yr dataset \citep{12p5yr_timing}.
        We compare the recovered values from each model to the results of \cite{12p5yr_gwb} with the dashed--dotted black line being the median value and the shaded pink region their 5\%--95\% quantiles at a reference frequency of $f=1~\mathrm{yr}^{-1}$.
        \label{fig:2043_fact_like}
        }
    \end{figure}
    
    The inclusion of this additional RN process allows us to extract any RN contribution at the spectral index predicted to be the most likely source of a stochastic gravitational wave background, which allows us to examine whether there is still evidence for RN in J2043+1711 in addition to the common spectral process found in the data set.
    We find that the $\gamma=13/3$ process corresponds to a median amplitude (with $1\sigma$ errors) of $\mathrm{log}_{10}(A)=-14.8^{+0.4}_{-1.8}$ with a SDBF of $2.2$ for model B and $\mathrm{log}_{10}(A)=-14.7^{+0.4}_{-2.0}$ and a SDBF of $2.0$ for model C.
    We show the results of this in figure \ref{fig:2043_fact_like}.
    Both are consistent with the results of the common spectral process found by \cite{12p5yr_gwb}.
    In addition, the intrinsic RN parameters still prefer a steep spectral process, albeit with a weak significance and with different median values for the two models: $\mathrm{log}_{10}(A) = -16.0^{+1.5}_{-2.7}, \gamma_{\mathrm{median}}=3.7^{+2.1}_{-2.4}, \mathrm{SDBF}=0.9$, and $\mathrm{log}_{10}(A)=-14.9^{+1.0}_{-3.2}, \gamma_{\mathrm{median}}=3.3^{+2.2}_{-2.0},  \mathrm{SDBF}=1.3$ for model B and C, respectively.
    Thus, using our methods, we have detected RN in PSR J2043+1711 for the first time.
    Furthermore, including the newly detected RN, we find significant changes in the timing model parameters.

\section{PSR J1600--3053}
\label{sec:J1600}

Pulsar J1600--3053 was discovered in the Parkes-Murriyang Telescope survey of high latitudes with a period of 3.6 milliseconds and in a binary orbit of 14.3 days \citep{Jacoby2007}.
A study of J1600--3053 in Parkes PTA (PPTA) data resulted in a significant detection of the Shapiro delay parameters $m_{\mathrm{p}} = 2.4(1.7)~\mathrm{M}_{\odot}$, $m_{\mathrm{c}} = 0.34(15)~\mathrm{M}_{\odot}$, and $\mathrm{sin}i = 0.87(6)$ \ark{($\mathrm{cos}i = 0.49\pm.01$)}, in addition to a measurement of $\dot{x} = -4.2(7) \times 10^{-15}$ \citep{Reardon2016}.
A separate study using European PTA (EPTA) data by \cite{Desvignes2016} measured the \ark{orthometric Shapiro delay parameters} $h_3 = 0.33(2)~\mu\mathrm{s}$ and $\varsigma = 0.68(5)$, and $\dot{x} = -2.8(5)\times10^{-15}$, all of which are consistent with the results of \cite{Reardon2016}. 
These \ark{orthometric Shapiro delay parameters} use a Fourier expansion of the system's \ark{orbital motion} to express the observed Shapiro delay \citep{Freire2010}.
The relevant post-Keplerian parameters, $h_{3}$ and $h_{4}$, are the Shapiro delay's third and fourth harmonic amplitudes.
The orthometric model is particularly useful in systems with low eccentricity, but is also more numerically stable for systems with strong Shapiro delay \citep{FonsecaThesis}.

In NANOGrav's 11-yr data, \cite{11yr_timing} made a significant detection of the Shapiro delay parameters, $\dot{x}$, and $\dot{\omega}$ to place a constraint of $m_{\mathrm{p}} = 2.4^{+0.7}_{\text{--}0.6}~\mathrm{M}_{\odot}$. However in their 12.5-yr data release, \cite{12p5yr_timing} found that $\dot{\omega}$ was no longer significant.
\cite{12p5yr_timing} did place another pulsar mass constraint on J1600--3053 of $m_{\mathrm{p}} = 2.0^{+1.1}_{\text{--}1.0}~\mathrm{M}_{\odot}$.
In the PPTA's second and most recent data release, \cite{Reardon2021} made a detection of $\dot{\omega}=0.0043(6)~\mathrm{deg}~\mathrm{yr}^{-1}$ and thus were able to put a tighter constraint on the pulsar's mass of $m_{\mathrm{p}} = 2.06^{+0.44}_{\text{--}0.41}~\mathrm{M}_{\odot}$.

\subsection{Noise Analysis}
\begin{figure*}[!htbp]
        \centering
        \includegraphics[width=\textwidth]{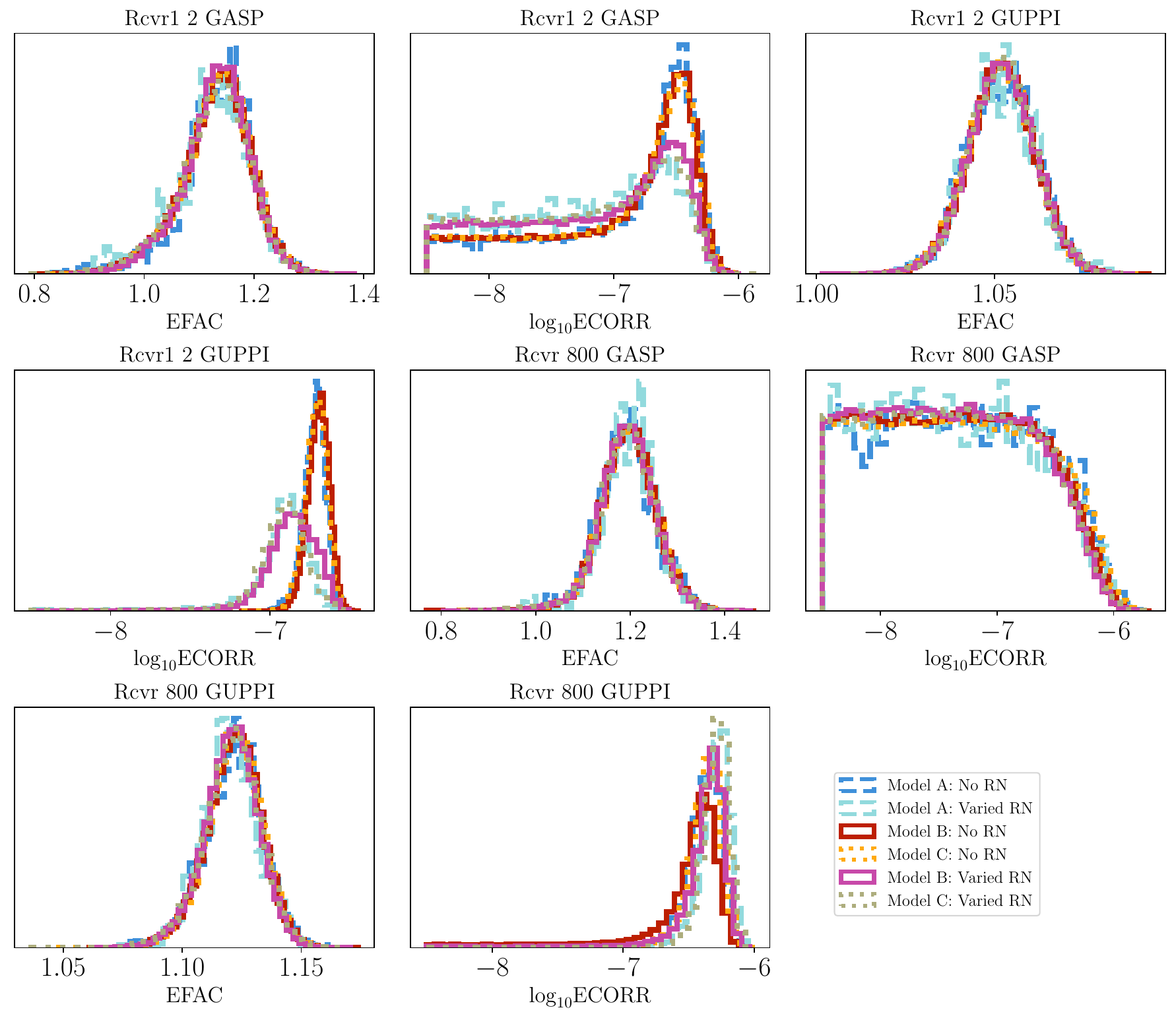}
        \caption{White noise parameter posteriors separated by observing backends for PSR J1600--3053 using the NANOGrav 12.5-yr dataset \citep{12p5yr_timing}.
        The histograms show model A with and without RN as the dashed cyan lines and blue lines, respectively, model B with and without RN as the solid red and solid magenta lines, respectively, and model C with and without RN as the dotted orange and dotted light olive--green lines, respectively.
        }
        \label{fig:J1600_Noise_Plots}
    \end{figure*}

    We perform the first fully general Bayesian analysis on J1600--3053 using the NANOGrav 12.5-yr data set.
    For our analysis of J1600--3053, we focus on the effect of directly sampling the timing model parameters with varying models for noise.
    In figure \ref{fig:J1600_Noise_Plots}, we compare the gamut of our models beginning with the most basic, and most often used, model in \texttt{ENTERPRISE} to analyze an individual pulsar: model A, where all timing model parameters are analytically-marginalized and linearized, and include intrinsic RN.
    Model A is the model used in the analysis of \cite{12p5yr_gwb} to search for a stochastic GWB, and thus is one of the most important to compare to since it reflects the noise levels used in a full PTA search for GWs. 
    When comparing model A to models B and C when intrinsic RN is included in the model, we find very little differences.
    Since the noise parameters are consistent with each other, we assume that the timing model behavior in the linearized model A should closely mirror that of the timing model posteriors of the same model, but with the same parameters directly sampled.
    In fact, we see that the MCMC sampled timing model posteriors between model B and C with RN behave almost identically despite model C marginalizing over DMX, jumps between telescope backends, and frequency dependent coefficient parameters. This follows from the fact that this last set of parameters are all linear components of the full timing model.
    We show this in table \ref{tab:J1600_tm_changes}.
    
\begin{figure*}[!htbp]
        \centering
        \includegraphics[width=\textwidth]{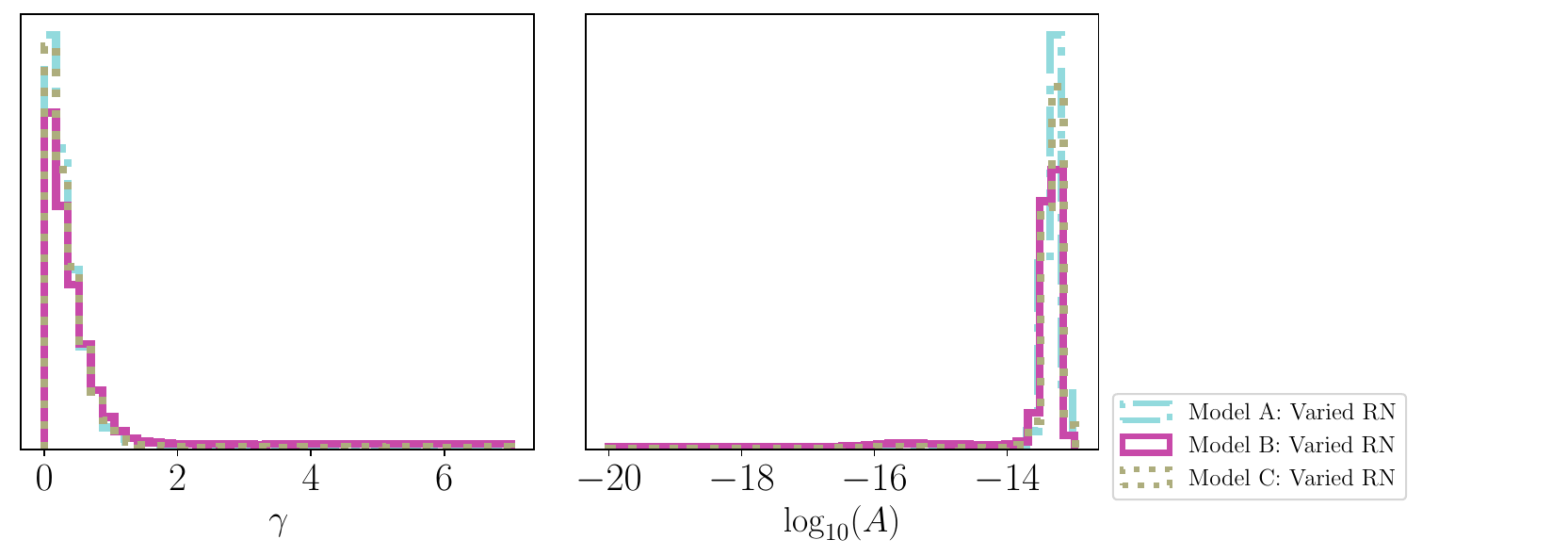}
        \caption{Red noise parameter posteriors PSR J1600--3053 using the NANOGrav 12.5-yr dataset \citep{12p5yr_timing}.
        The histograms show model A as the dashed cyan lines, model B as the solid magenta lines, and model C as the dotted light olive--green lines.
        }
        \label{fig:J1600_Red_Noise_Plots}
    \end{figure*}

    In all the analyses with intrinsic RN included, shown in figure \ref{fig:J1600_Red_Noise_Plots}, we find that the spectral index in each prefers a flat spectrum ($\gamma=0$), which behaves like white noise, with a fairly large RN amplitude $-14<\mathrm{log}_{10}(A)<-13$.
    We estimate the significance of each with a SDBF and find a SDBF of 5.4(3) for model B, 13.0(5) for model C, and that the model A RN is separated from the lower prior bound thus giving an infinite SDBF since the approximation breaks down.
    Since the RN parameters prefer a more white noise-like spectrum with some significant evidence, there is clearly excess power not accounted for in any of the models.
    Because the SDBF decreases between model A and model C, the explicitly varied timing model accounts for some of the excess noise.
    As the SDBF further decreases between model C and model B, we suspect the numerically marginalized ``nuisance parameters" better account for the unmodeled noise.
    The interplay between noise and timing model parameters requires a more in-depth study that falls outside of the scope of this paper.

    Without including intrinsic RN, we find that the noise parameters shift to higher, more constrained values of ECORR in all backends.
    Since we are removing the intrinsic RN, a significant portion of the noise modeling, we expect the WN to increase.
    Since previous noise analyses find no evidence for RN and thus do not include it in their modeling, the analysis without RN is more consistent with \cite{12p5yr_timing} for comparison.

    We find that for all timing model parameters for both model B and C are within the one sigma region of the published GLS values.
    For all parameters except the spin frequency, model B improves upon the one-sigma error constraints of the GLS analysis.
    Model C decreases the one-sigma error region as compared to the GLS analysis for the position parameters, $\lambda$ and $\beta$, as well as the proper motion in each position, the binary period, the decay in semi-major axis, as well as the eccentricity, incination angle, and consequently the pulsar mass.

\subsection{Pulsar Mass}
\begin{figure*}[!htbp]
        \centering
        \includegraphics[width=\textwidth]{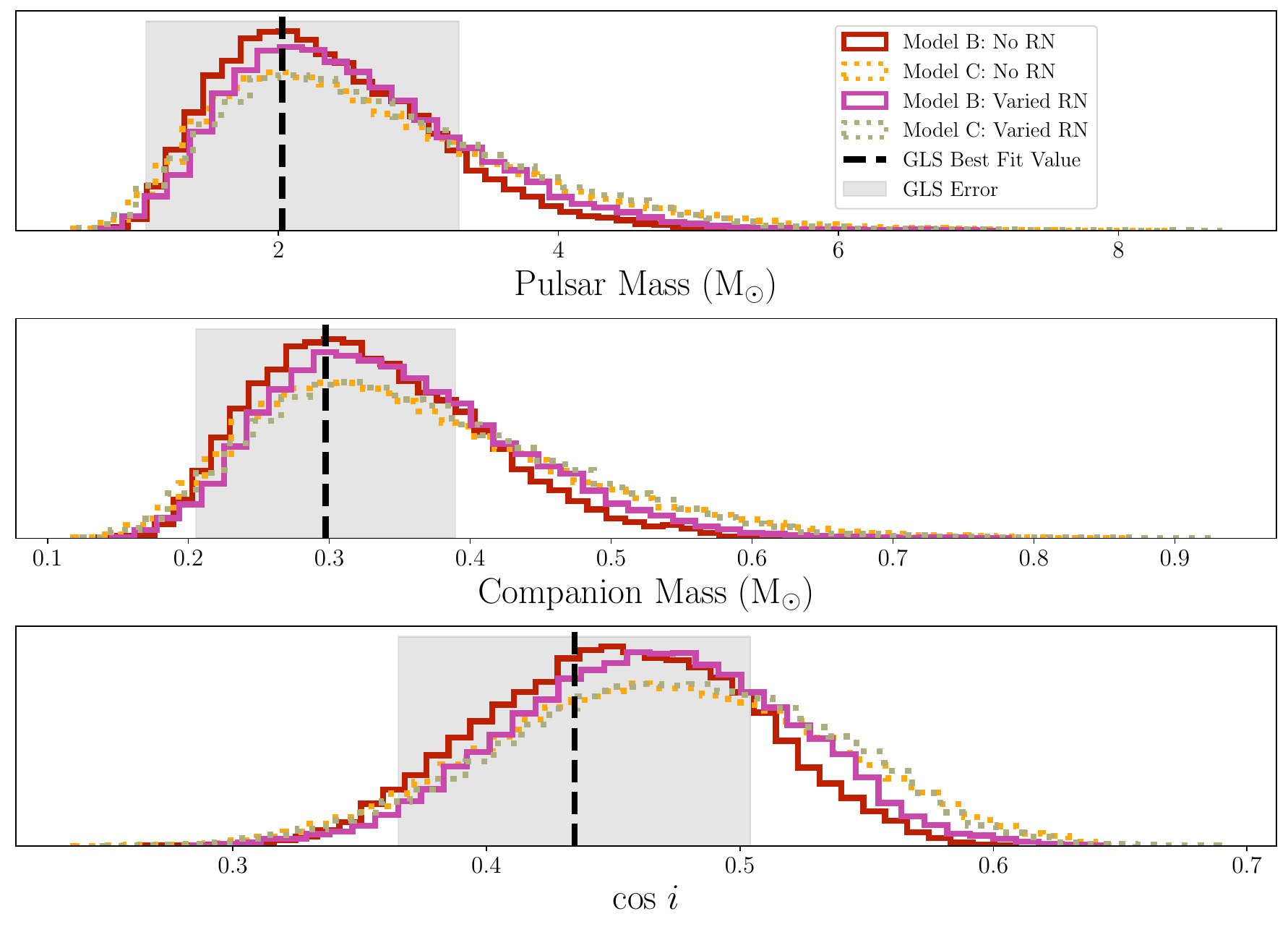}
        \caption{Shapiro delay parameter plots for PSR J1600--3053 using the NANOGrav 12.5-yr dataset \citep{12p5yr_timing}.
        The histograms show model B with and without RN as the solid red and solid magenta lines, respectively, and model C with and without RN as the dotted orange and dotted light olive--green lines, respectively.
        The dashed black line represents the best fit value from the GLS analysis of \cite{12p5yr_timing} with the shaded region being the respective one-sigma error regions.
        }
        \label{fig:J1600_Mass_Plots}
    \end{figure*}

    Despite the difference between the various model B and C with and without RN, we find \ark{that the} mass parameters for J1600--3053 are all consistent with the GLS best-fit value.
    We show the posteriors for the inclination of the system ($\mathrm{cos}i$), along with the companion mass ($m_{\mathrm{c}}$) and the inferred pulsar mass ($m_{\mathrm{p}}$) based on the mass function (see equation \ref{eq:mass_function}) in figure \ref{fig:J1600_Mass_Plots}.
    
    While the posteriors are still consistent with the GLS timing best-fit parameters, there is still some discrepancy with respect to the center of the distributions.
    We find that the cosine of the inclination angle shifts to higher values for each model.
    The model B and C parameters for $\mathrm{cos}i$, $m_{\mathrm{c}}$, and consequently $m_{\mathrm{p}}$ broadly follow the same distributions regardless of whether RN is included in the model.
    We do see improved constraints in each model B posterior when compared to the model C one-sigma region.
    We report median values and their $68\%$ credible intervals (CIs) for model B and C without RN in table \ref{tab:J1600_tm_changes}.
    We do not report model B and C with RN as the resulting values without RN are identical.

\section{PSR J0740+6620}
\label{sec:J0740}

Pulsar J0740+6620 was initially discovered in the Green Bank Northern Celestial Cap Pulsar Survey with a period of $2.89$ milliseconds and a DM of $14.96(4)$ \citep{GBNCCI}.
Later in \cite{GBNCCIII}, a full timing solution was presented and J0740+6620 was found to have a \ark{companion object}. 
PSR J0740+6620 is in a nearly circular orbit ($e=5\times10^{-6}$) with a period of 4.77 days.
Searches by \cite{Beronya2019} in optical and near-infrared found the likely companion to be an ultra-cool helium-atmosphere white dwarf (WD).
\cite{Cromartie2020}, using concentrated observing campaigns during the binary system's superior conjunction, in addition to observational data from \cite{12p5yr_timing}, estimated J0740+6620 had the largest mass of a neutron star to date of $2.14^{+0.10}_{\text{--0.09}}$ at a CI of 68.3\%.
\cite{Fonseca2021} combined data from \cite{Cromartie2020}, \cite{12p5yr_timing}, and high cadence observations from the Canadian Hydrogen Intensity Mapping Experiment (CHIME) to set a new mass estimate on J0740+6620 of $m_{\mathrm{p}}=2.08^{+0.07}_{\text{--}0.07}$ at a CI of 68.3\%.
In addition to the new mass constraints, \cite{Fonseca2021} measured secular variations in the orbital period of the binary and set a new distance constraint from timing, indicating the \ark{companion object} is truly an ultra-cool WD.

We perform analyses on three data sets: those of \cite{12p5yr_timing}, \cite{Cromartie2020}, and \cite{Fonseca2021}.
Each data set uses the data from the previous data sets and extends them in unique ways such that \cite{12p5yr_timing} $\subset$ \cite{Cromartie2020} $\subset$ \cite{Fonseca2021}.
Thus we not only explore the properties of each individual data set with our fully general method, but also trace changes across time and additional observing programs.
In \cite{Cromartie2020}, the authors found that J0740+6620's DM varied smoothly over the course of the data set, but \ark{found that} the DM behavior was not fully characterized by a quadratic fit using only the first two time derivatives of DM.
With their extended data set, the authors of \cite{Fonseca2021} used several DMX-binning schemes to account for systematic uncertainties associated with a choice in DMX modeling.
They separated the full data into three sets of timing models with DMX binned by 3 and 6.5 days, and a hybrid model blending both over the low and high cadence observation periods.
In this work, we use the data set produced by binning the DMX every 3 days as that was the nearest to the model-averaged results when using the data from \cite{Fonseca2021}. 

\subsection{Timing Model Analysis}
    Overall, we find extremely good agreement between the different models for the individual data sets.
    For the NANOGrav 12.5-yr data set, out of both model B and C analyses, only the ecliptic latitude position parameter for model B has a median posterior outside of the the GLS best-fit one-sigma region.
    The same model B analysis improves upon the one-sigma constraints set by the GLS analysis for several of the parameters, including the parallax, binary period, and inclination angle.
    The model B and C analyses on the \cite{Cromartie2020} data have similar results; only the ecliptic latitude parameter median lies outside the previously reported one-sigma region.
    For model B, tighter constraints are set on more parameters, including the companion mass, inclination angle, the position parameters, and others.
    The model B analysis on the data from \cite{Fonseca2021} improves upon all parameters.
    There are shifts in both model B and C in the ecliptic latitude parameter, and the median posterior values are nearly two-sigma away.
    For a complete summary of the timing model posterior changes from these analyses, see tables \ref{tab:J0740_tm_changes_pt1}, \ref{tab:J0740_tm_changes_pt2}, and \ref{tab:J0740_tm_changes_pt3}.

\subsection{Noise Analysis}
    We find that all the white noise parameters and the timing model parameter posteriors are statistically indistinguishable whether or not RN parameters are included both for the NANOGrav 12.5-yr data set and that of \cite{Cromartie2020} for both model B and C.
    In both data sets and models, the RN paramater posteriors return the uniform prior for the spectral index and show no evidence for RN ($\mathrm{SDBF} < 1$).
    For the same analyses with the data from \cite{Fonseca2016}, the white noise parameters again show very little to no difference with or without the RN parameters included.
    The RN posteriors for model B behave similarly to the other two data sets, but the model C analyses slightly prefer a flat spectral index around $\gamma=0$, but with very little evidence in the amplitude parameter ($\mathrm{SDBF} < 1$).
    The timing model parameters do show a slight broadening when RN is included, but the medians of their posteriors are well within the one-sigma region of the GLS best-fit estimations.

\subsection{Pulsar Mass}
\begin{figure*}[!htbp]
        \centering
        \includegraphics[width=\textwidth]{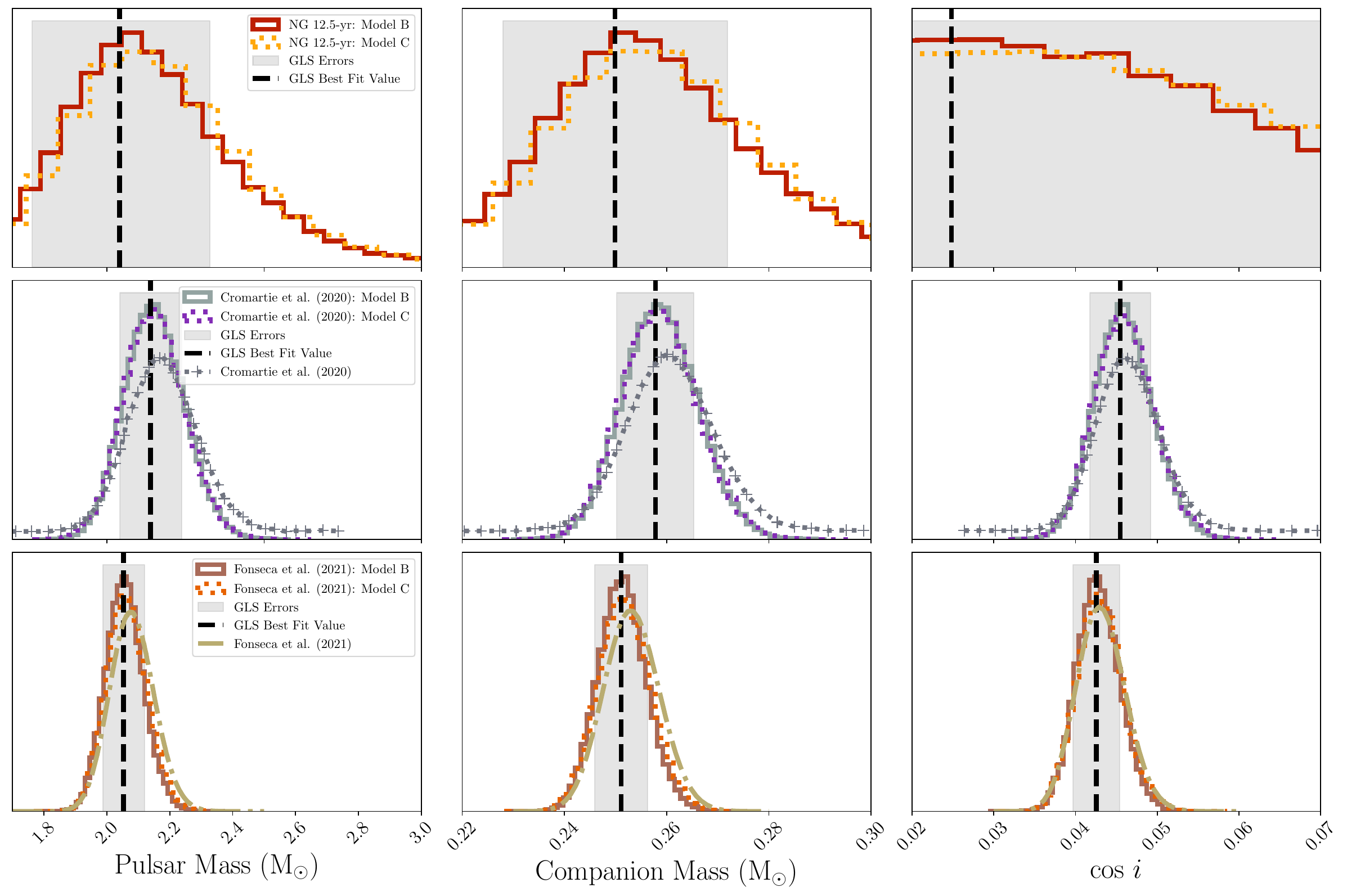}
        \caption{Shapiro delay parameter plots for PSR J0740+6620. 
        The top three panels contain the results of a model B (solid red), and a model C (dashed orange) on the NANOGrav 12.5-yr dataset \citep{12p5yr_timing}. 
        The middle three panels contain the results for the analyses using a model B (solid light gray), a model C (dashed purple) analysis, as compared to the results from \cite{Cromartie2020} (dotted--crossed bluish--gray line).
        The bottom three panels contain the results for the analyses using a model B (solid brown), a model C (dashed red--orange) analysis, as compared to the results from \cite{Fonseca2021} (dashed--dotted olive--green line).
        The dashed black line represents the best fit value from the respective GLS analysis of each dataset with the shaded region being their respective one-sigma error regions. 
        None of the represented analyses include RN.
        }
        \label{fig:J0740_Mass_Plots}
    \end{figure*}

    As seen in figure \ref{fig:J0740_Mass_Plots}, the concentrated observations of J0740+6620 during superior conjunction by \cite{Cromartie2020} allow for greatly improved constraints on binary's inclination and component masses.
    Likewise, the near-daily cadence observations of \cite{Fonseca2021} improve the constraints further.
    We find that our analyses have median values consistent with each of the data set's reported best-fit values and constraints for both model B and C.
    Furthermore, our analyses are consistent with the chi-square gridding techniques used in both \cite{Cromartie2020} and \cite{Fonseca2021}, but with better constraints and slightly lower values for the companion mass and subsequently the derived pulsar mass.
    For our analysis of the NANOGrav 12.5-yr data set from \cite{12p5yr_timing}, the companion mass is $0.26(2)$ and $\mathrm{cos}i=0.04(4)$ for both models, and thus the derived pulsar mass is $m_{\mathrm{p}}=2.1^{+0.3}_{-0.2}$ for models B and C.
    The analysis of the data set from \cite{Cromartie2020}, the companion mass is $0.259(8)$ and $\mathrm{cos}i$ is $0.046(4)$ for both model B and C, and thus the derived pulsar mass is $m_{\mathrm{p}}=2.15^{+0.1}_{-0.09}$ for model B and C.
    The analysis of the data set from \cite{Fonseca2021}, the companion mass is $0.251(5)$ and $0.252(5)$ for model B and C, respectively, while $\mathrm{cos}i$ is $0.043(3)$ for both models, and thus the derived pulsar mass is $m_{\mathrm{p}}=2.06(6)$ for model B and $m_{\mathrm{p}}=2.06^{+0.07}_{-0.06}$ for model C, both of which improve on the constraints set by the GLS fit and the gridded chi-square fit of \cite{Fonseca2021}.

\section{PSR J1640+2224}
\label{sec:J1640}

Pulsar J1640+2224 was discovered in an Arecibo survey of high Galactic latitudes \citep{Foster1995A,Foster1995B} with a period of 3.16 milliseconds and an orbital period of 175 days with a binary companion of estimated mass $m_{\mathrm{c}}=0.31~\mathrm{M}_{\odot}$.
Not long after, \cite{Lundgren1996} detected the companion with the Palomar 200-inch telescope, and based on its luminosity, distance, temperature, and \ark{age,} inferred that it was a helium WD.
Using combined data from the Effelsberg 100--m and Arecibo 305--m radio telescopes, \cite{Lohmer2005} detected the effects of Shapiro delay and placed tight constraints on the binary companion's mass and system inclination $m_{\mathrm{c}}=0.17^{+0.08}_{\text{--}0.05}~\mathrm{M}_{\odot}$ and $\mathrm{cos}i=0.11^{+0.09}_{\text{--}0.07}$, respectively, confirming it as a WD companion.
The timing parameters for J1640+2224 continued to improve with more observations.
\cite{Vigeland2014}, using data from the NANOGrav 5-yr data set \citep{Demorest2013}, used Bayesian inference and linearized least-squares techniques to compare each method's timing solutions for J1640+2224.
They found that using Bayesian modeling recovers slightly different timing parameters, especially when placing priors on the companion mass centered around the recovered WD mass of \cite{Lundgren1996}.
The first European PTA data release found a significant detection of the derivative in projected semi-major axis over time of $\dot{x}=1.07(16)\times10^{-14}$ \citep{Desvignes2016}, but failed to detect significant Shapiro delay or parallax.
The $\chi^{2}$-gridding technique developed in \cite{Splaver2002} was employed by \cite{Fonseca2016} on J1640+2224 and made detections of $\dot{x}=1.45(10)\times10^{-14}$, change in longitude of periastron $\dot{\omega}=\text{--}2.8(5)\times10^{\text{--}4}~\mathrm{deg}~\mathrm{yr}^{-1}$, and the Shapiro delay parameters: companion mass $m_{\mathrm{c}}=0.6^{+0.04}_{\text{--}0.02}~\mathrm{M}_{\odot}$, inclination angle $\mathrm{cos}i=60(6)$, and pulsar mass $m_{\mathrm{p}}=4.4^{+2.9}_{\text{--}2.0}~\mathrm{M}_{\odot}$.
As a follow-up on the previous optical and radio observations, \cite{Vigeland2018} combine data from the Very Long Baseline Array (VLBA) and a reanalysis of archival optical observations with the Wide Field Planetary Camera 2 (WFPC2) on the Hubble Space Telescope (HST) to attempt to constrain the WD companion mass and its type (either a low-mass helium-core WD or a high-mass carbon-oxygen (CO) WD).
They set a lower limit on the WD mass of $0.4~\mathrm{M}_{\odot}$ and conclude the companion is most-likely ($>90\%$ confidence) a CO WD.

In this work, we analyze pulsar J1640+2224 in the data sets from NANOGrav's 5-yr, 9-yr, and 12.5-yr \citep{Demorest2013,9yr_timing,12p5yr_timing}.
We examine the effects in each of these analyses by setting a limit on the maximum allowable mass for the pulsar (determined by equation \ref{eq:mass_function}) of $m_{\mathrm{p}}<3~\mathrm{M}_{\odot}$.
We choose PSR J1640+2224 in particular because analyses of its mass has consistently been greater than the predicted maximum for neutron stars.

PSR J1640+2224 has been shown in several analyses to be difficult to model \citep{Vigeland2018, Fonseca2016}. In each of the datasets we examined, the convergence criteria set in section \S\ref{subsec:global_model_params} of a Gelman-Rubin split R-hat statistic for each parameter of $<1.1$ and auto-correlation lengths of $<200$ were difficult to meet and often required much longer run times. Our model C analyses for the NANOGrav 5-yr and 12.5-yr datasets fully converge, while the model B analyses do not conform to our convergence standards for each analysis with both datasets.
The model B and model C analyses for the NANOGrav 9-yr dataset both have a Gelman-Rubin score of $<1.1$ for all parameters, but have auto-correlation lengths of $>200$. This indicates convergence, but without many significant samples.

We expect that with enough samples, all non-converged analyses would eventually converge, but due to computational constraints we are unable to achieve full convergence here.

We use our analysis of the NANOGrav 5-yr data set to compare to the results of \cite{Vigeland2014} and their use of a mixed linearized/full timing model, much like our model C analysis.
We also compare our analysis to those of \cite{Fonseca2016} for J1640+2224 using the NANOGrav 9-yr data set, where they used the chi-squared gridding techniques to place constraints on the component masses and system inclination.
Finally, we use the NANOGrav 12.5-yr data set to examine J1640+2224 in depth with Bayesian timing for the first time and compare the results of our model B and C analyses with and without restrictions on the pulsar mass.

\subsection{NANOGrav 5-yr Dataset}
    We analyze the NANOGrav 5-yr dataset from \cite{Demorest2013} using our model B and C analyses with and without RN.
    While the model C analysis conforms to our criteria of convergence, after over 7.3 million MCMC steps we were only able to reach a Gelman-Rubin split R-hat statistic for the least converged variable of 1.2 with an autocorrelation length of $2.0\times10^{5}$ for the model B analysis. Again we expect that with enough samples, the model B analyses would converge, but due to computational constraints we are unable to achieve full convergence here.

\begin{figure*}[!htbp]
        \centering
        \includegraphics[width=\textwidth]{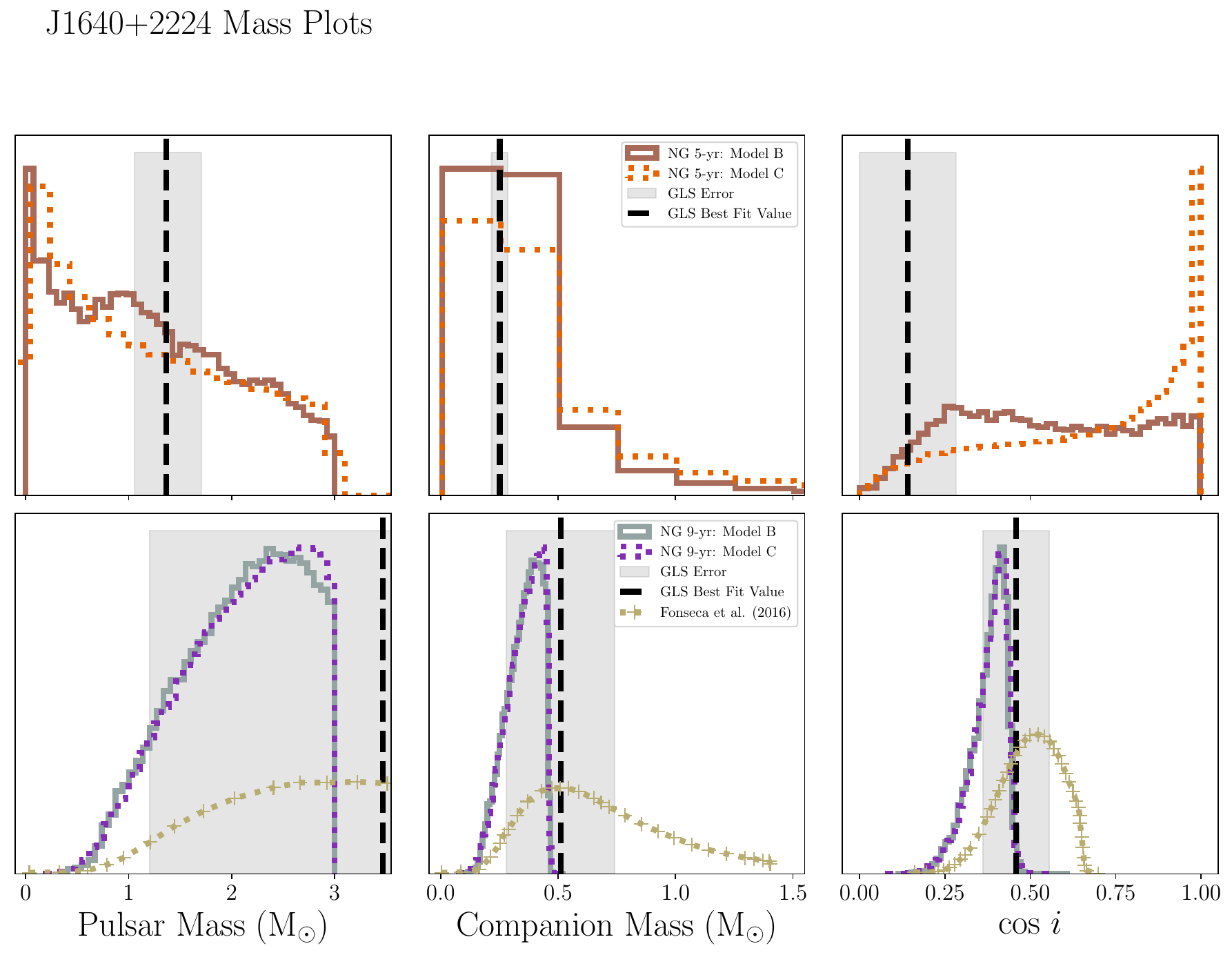}
        \caption{Shapiro delay parameter plots for PSR J1640+2224. 
        The top three panels contain the results of a model B (solid brown), and a model C (dashed red--orange) on the NANOGrav 5-yr dataset \cite{Demorest2013}. 
        The bottom three panels contain the results for the NANOGrav 9-yr dataset \citep{9yr_timing} using a model B (solid light--gray), and a model C (dashed purple) analysis compared to the chi-squared gridding results of \cite{Fonseca2016} (dotted-crossed olive--green line). 
        The dashed black line represents the best fit value from the GLS analysis of each dataset (the top being \cite{Demorest2013} and the bottom \cite{9yr_timing}) with the shaded region being their respective one-sigma error regions.
        RN is not included in these models.
        }
        \label{fig:J1640_5yr_9yr_Mass_Plots}
    \end{figure*}

    We find no evidence that there is RN in the data.
    In both models, when RN was included the posteriors for the RN spectral index returned the uniform prior and all SDBFs for the RN amplitude were less than 1.
    We found also that there was no difference between WN parameters between model B and C, nor when RN was included.
    We do find discrepancies between the median posterior values and their 68\% CI for the majority of timing model parameters when using either model as compared to the GLS results published in \cite{Demorest2013} (see table \ref{tab:J1640_tm_changes_5yr}).
    Using this dataset, we find little evidence for the inclusion of Shapiro delay.
    The data, when examined with our Bayesian methods finds a companion mass consistent with zero, a relatively unconstrained inclination angle, and consequently, a pulsar mass consistent with zero.
    These findings are shown in figure \ref{fig:J1640_5yr_9yr_Mass_Plots}.
    
    We find that despite using the same priors as in \cite{Vigeland2014}, we obtain significantly different results for some parameters.
    As mentioned in section \S\ref{subsec:global_model_params}, we use an informative prior of the form of equation \ref{eq:DM_prior} on the parallax based on the distance estimation from J1640+2224's DM as in \cite{Vigeland2014}.

    We find that there is not much evidence for the inclusion of parallax with the limited observation timespan of NANOGrav's 5-yr dataset as the posterior of the parallax is the same as the input prior.
    We find a significant difference in the posterior distribution of \cite{Vigeland2014} for the projected semimajor axis, $x$; where the results of \cite{Vigeland2014} appear broad and one-sided towards values lower than the original GLS timing model fit, our results, while also broad, are Gaussian and centered around the published value of the NANOGrav 5-yr value.
    Our results are not wholly different from those of \cite{Vigeland2014}, as our posteriors for companion mass and inclination angle closely match their posteriors.
    Much of the parameter differences are that our results are much less constraining than those of \cite{Vigeland2014}, but both are generally centered around the GLS timing model fit results.
    
\subsection{NANOGrav 9-yr Dataset}
    We now analyze the NANOGrav 9-yr dataset from \cite{9yr_timing} using our model B and C analyses with and without RN.
    Unlike our analyses on the NANOGrav 5-yr dataset, the model B and model C analyses both achieve a Gelman-Rubin score of less than 1.1, but the long autocorrelation lengths for the spin frequency ($\nu$) result in an effective sample size of 219 and 31 for model B and model C, respectively. So while the analyses are converged based on the Gelman-Rubin statistic, the effective sample size remains low which can affect the significance of the resulting distributions.
    We again find no evidence for RN: the spectral index posteriors are broad, but with a slight preference for a spectral index of zero, and the SDBF for the RN amplitude is much less than one.
    We find little to no difference between the model B and C WN posteriors, with or without RN.
    As opposed to our analysis of NANOgrav's 5-yr dataset, the timing model posteriors are, with the exception of a few parameters shown in table \ref{tab:J1640_tm_changes_9yr}, consistent with the best-fit values and errors of \cite{9yr_timing}.

    The most evident changes occur in some of the Keplerian parameters and the parallax parameter.
    We find that the projected semimajor axis ($x$), the cosine of the inclination angle, the companion mass, and the parallax all result in very asymmetric posteriors. 
    The parallax parameter has a median value against the bound of the physical prior we impose, which indicates the lack of evidence for the inclusion of parallax since negative values of parallax are unphysical.
    The asymmetry in $x$, $\mathrm{cos}i$, and $m_{\mathrm{c}}$ could indicate these parameters are non-linear, but more likely this is an effect of only allowing pulsar masses less than three solar masses. 
    Since $m_{\mathrm{p}}$ is dependent on $x$, $P_{\mathrm{b}}$, $\mathrm{sin}i$, and $m_{\mathrm{c}}$, according to equation \ref{eq:mass_function}, by restricting the pulsar mass, we are necessarily limiting the combined space each of the determining parameters explore. 
    As one can see in figure \ref{fig:J1640_5yr_9yr_Mass_Plots}, with the imposed bound on pulsar mass, the binary mass parameters and the inclination angle appear to all have a strong boundary because of the pulsar mass bound.
    The mass bound causes tension between our results and those of \cite{Fonseca2016} as well as from \cite{9yr_timing}.
    With the restrictions on the pulsar mass, we find the companion mass to be $m_{\mathrm{c}}=0.36(9)$ and $m_{\mathrm{c}}=0.37(9)$, for model B and C, respectively, while $\mathrm{cos}i$ is $\mathrm{cos}i=0.39(6)$ for both models. Thus the derived pulsar mass is $m_{\mathrm{p}}=2.1^{+0.6}_{-0.7}$ for model B and $m_{\mathrm{p}}=2.2^{+0.6}_{-0.7}$ for model C.
    In section \S\ref{subsec:J1640_12p5yr}, we explore the data of \cite{12p5yr_timing} with and without the bound on pulsar mass.

\subsection{NANOGrav 12.5-yr Dataset}
    \label{subsec:J1640_12p5yr}

    We finally analyze J1640+2224 with the NANOGrav 12.5-yr dataset of \cite{12p5yr_timing}.
    Similarly to the analyses for the NANOGrav 5-yr dataset, the model C analyses are fully converged according to our standards in section \S\ref{subsec:global_model_params} for both the restricted and unrestriced mass analyses. The model B analyses however, were only able to reach a Gelman-Rubin split R-hat statistic for the least converged variable of 5.2 and 6.0 and an autocorrelation length of $2.0\times10^{5}$ and $1.8\times10^{5}$ in the restricted and unrestricted model B analyses, respectively, after over 1.6 million MCMC steps. Again we expect that with enough samples, the model B analyses would converge, but due to computational constraints we are unable to achieve full convergence here. Thus the model B results may not represent the true distribution of each parameter.
    We compare our model B and C analyses with and without RN and find there is no evidence for RN in J1640+2224.
    While there is some discrepancy in the RN amplitude posteriors between model B and C, both sets of posteriors return $\mathrm{SDBF}<1$ and posteriors that fill the prior bounds on the RN spectral index.
    The WN parameters do not depend on whether RN is included, but the posteriors for all ECORR parameters have separated median values, or even have completely different distributions between model B and C.
    We find that this is true regardless of whether we restrict the pulsar mass.

    \begin{figure*}[!htbp]
        \centering
        \includegraphics[width=\textwidth]{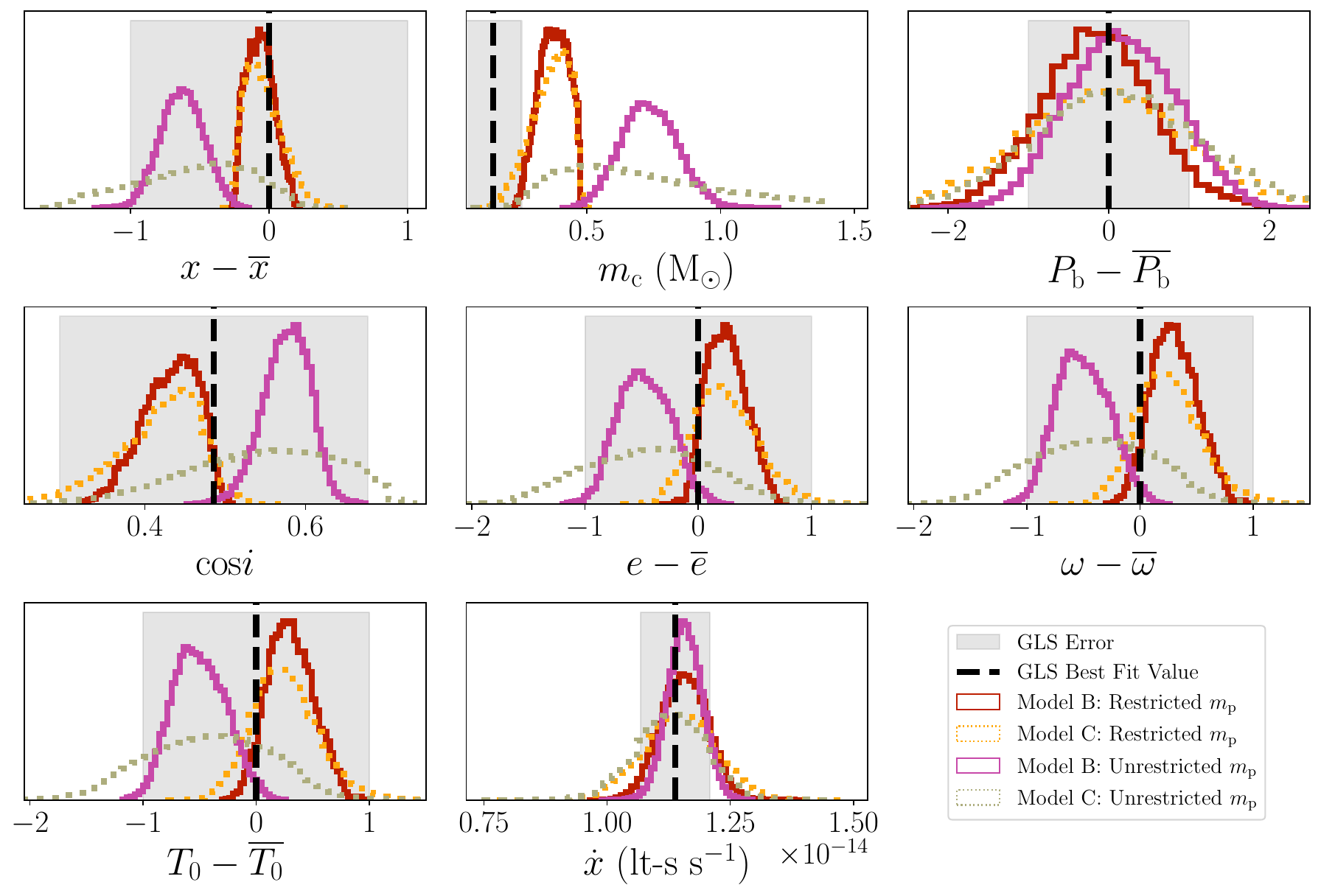}
        \caption{Comparison of Keplerian parameters of PSR J1640+2224 for model B with and without restrictions on the pulsar mass, as shown in the solid magenta and solid red histograms, respectively and model C with and without restrictions on the pulsar mass, as shown in the dotted light olive--green and dotted orange histograms, respectively.
        Each analysis is done on the NANOGrav 12.5-yr dataset \citep{12p5yr_timing} without the inclusion of RN.
        The dashed black line represents the best fit value from the GLS analysis of \cite{12p5yr_timing} with the shaded region being the respective one-sigma error regions.
        }
        \label{fig:J1640_Keplerian_Plots}
    \end{figure*}

    As is expected, the largest differences between the analyses that restrict and do not restrict the pulsar mass occur in the Keplerian and post-Keplerian parameters.
    We show the most egregious discrepancies in figure \ref{fig:J1640_Keplerian_Plots}.
    When we restrict the pulsar mass, the model B and C posteriors have statistically similar distributions, but without the restriction, the model C posteriors are much broader than those of model B.
    Furthermore, the unrestricted analyses result in lower median values for inclination angle, eccentricity, longitude of periastron, and epoch of periastron.
    The lower values are still well within the error region of the GLS best-fit values, and the posterior distributions are often fully within the error regions.
    For both the restricted and unrestricted mass analyses, nearly all of the posteriors are within the one-sigma error region of the GLS analyses except the computed pulsar mass and the position parameters, $\lambda$ and $\beta$.

    \begin{figure*}[!htbp]
        \centering
        \includegraphics[width=\textwidth]{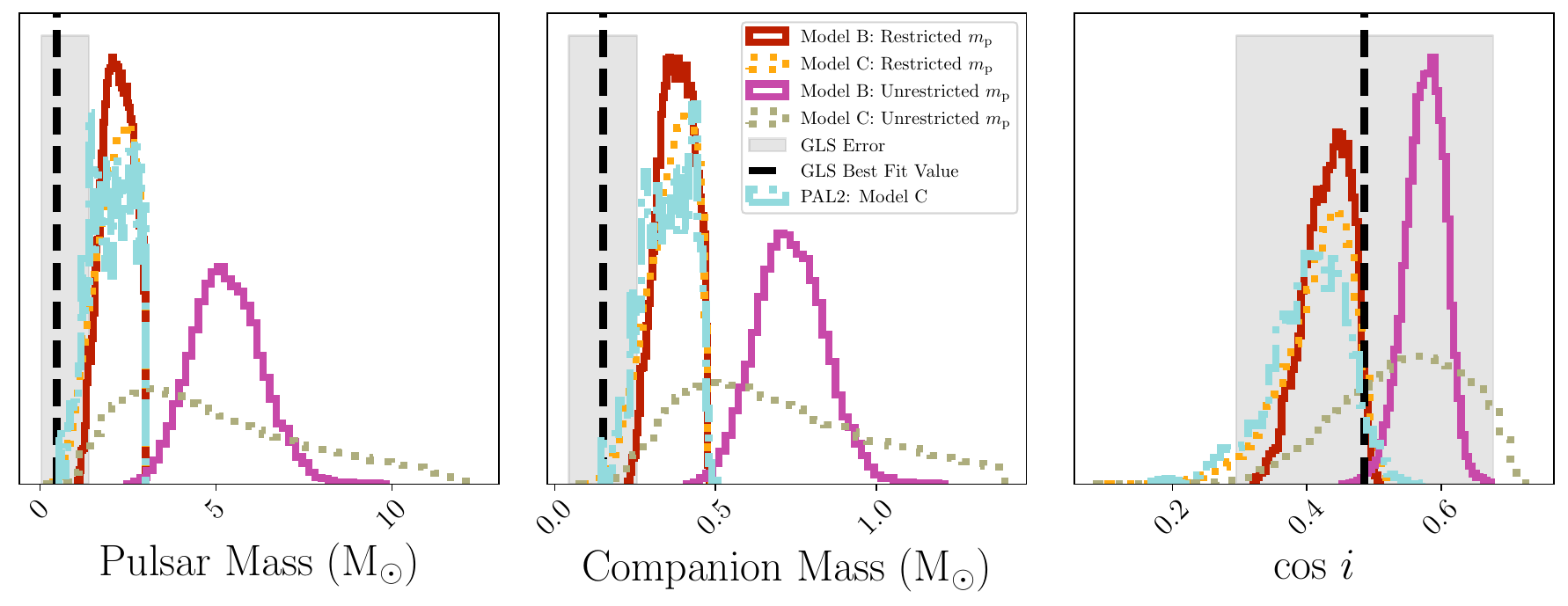}
        \caption{Shapiro delay parameter plots for PSR J1640+2224 using the NANOGrav 12.5-yr dataset \citep{12p5yr_timing} without the inclusion of RN. The histograms for model B are shown with and without restrictions on the pulsar mass by the solid magenta and solid red histograms, respectively and for model C with and without restrictions on the pulsar mass, as shown in the dotted light olive--green and dotted orange histograms, respectively.
        The dash--dotted cyan histogram shows the results for model C using PAL2.
        The dashed black line represents the best fit value from the GLS analysis of \cite{12p5yr_timing} with the shaded region being the respective one-sigma error regions.
        }
        \label{fig:J1640_12p5yr_Mass_Plots}
    \end{figure*}

    We show the results of model B and C analyses with and without the $3~\mathrm{M}_{\odot}$ restriction on the pulsar mass in figure \ref{fig:J1640_12p5yr_Mass_Plots}.
    Because of the shift to higher values for the companion mass and inclination angle for the unrestricted case study, the resulting pulsar mass has a median value of $m_{\mathrm{p}}=5.3^{+1.0}_{-0.9}$ and $m_{\mathrm{p}}=5^{+3}_{-2}$, respectively.
    For the model C analyses the one sigma region includes values below our assumed physical limit of $3~\mathrm{M}_{\odot}$, but the model B analyses does not include any values below $3~\mathrm{M}_{\odot}$ even at the two sigma level.
    In the restricted mass case, both model B and C analyses, by construction, have medians that fall below the $3~\mathrm{M}_{\odot}$ cutoff: $m_{\mathrm{p}}=2.2^{+0.5}_{-0.5}$ and $m_{\mathrm{p}}=2.2^{+0.5}_{-0.7}$, respectively.
    Each of their one sigma regions however include the boundary and thus likely influence the likelihood to stay at lower values.
    Both the restricted and unrestricted analyses have companion masses that are consistent within their one sigma region with the upper limit of \cite{Vigeland2018}, but only the unrestricted posteriors are consistent with the estimate by \cite{Fonseca2016} using the NANOGrav 9-yr data.
    While this leaves, in both cases, the pulsar mass higher than currently predicted by neutron star equation of state theories, the longer J1640+2224 is observed and studied, the tighter the constraints become on each of its Keplerian parameters. 

\section{Discussion and Conclusions}
\label{sec:Discussion}

Overall, we find our methods of generalized Bayesian timing with contemporaneous noise modeling to be consistent with other methods of timing including GLS timing, mixed analytically-marginalized linear/Bayesian timing such as \texttt{PAL2}, and chi-squared gridding techniques.
In most cases, we find that median posterior parameters align with the errors of these other timing methods and have similar confidence regions, many of which have improved constraints on timing parameters.
This trend appears most strongly when using the appropriate noise characterization (i.~e. whether to include RN) with the fully general model B.
\ark{The caveat to this conclusion being the poor convergence in the case of the model B analysis for PSR J1640+2224 with the NANOGrav 12.5 year data. See the discussion in Appendix \ref{sec:appendixB}}
The mixed model C often has broader posteriors when compared to the full model, but we find that it often converges much faster, on the order of twice as quickly.
This is particularly important when considering whether one should directly directly sample particular timing model parameters or analytically marginalize over them; there is a computation cost versus accuracy/covariance trade off when deciding between the two models.

\subsection{Application to GW Analyses}
    Another important aspect of the decision on whether to use the analytically-marginalized linear model or not is its behavior in the prior space and effect on the overall residuals.
    Given that the central assumption for the linear approximation used in marginalization is the Gaussianity of the likelihood in the relevant parameters, non-Gaussian behavior can indicate non-linear behavior.
    We find that most of our analyses result in Gaussian posteriors.
    Often non-Gaussian behavior occurs in parameters where we impose physical bounds on the prior such as parallax, binary component masses, and system inclination.
    
    In searches for GWs, each pulsar included in the full PTA search are evaluated based on how much support it has for a common-signal, such as a gravitational wave \citep{11yr_gwb,12p5yr_gwb}.
    As we have shown for pulsar J2043+1711, timing model parameters can exchange power with RN parameters.
    When evaluating these pulsars, one should consider whether the individual pulsar shows evidence of intrinsic RN as including RN when it is not present in the data can lead to shifts in timing model parameters.
    Furthermore, directly sampling the timing model parameters that are most likely to be affected by the inclusion of a common RN process can lead to improvement in the pulsar's contribution to the common signal.

\subsection{Implications for Pulsar Masses}
    We find that pulsar mass is for the most part consistent with values found in previous analyses for each pulsar.
    Our analyses, including full Bayesian timing models for all or some of the timing parameters, lead to tighter constraints on the pulsar mass than other rigorous studies using the latest data sets.
    As can be expected, with more data, especially around superior conjunction when the Shapiro delay is the strongest, the constraints on the binary component masses and inclination angle improve.
    We also find that each data set is unique and consequently data sets with longer time spans may not simply improve constraints around previous best-fit, median values, but they also can vary up to several sigma away from the most likely values using less data.
    We speculate that with enough data, this will no longer be the case.
    Eventually, we suspect, the timing model will become accurate and consistent enough that adding a few more years of observations will no longer shift parameters around, but simply improve their constraints.

\subsection{Future Use}
    The work here is endlessly extensible.
    The most immediate application, since these techniques rely upon initial GLS timing to create the initial timing model solutions, is using our methods as a final check on the behavior of the timing model in the presence of noise.
    Typically, once the timing models are constructed via GLS fitting, the solutions are used in an \texttt{Enterprise} individual pulsar noise analysis where the pulsar is assessed for RN and the maximum likelihood values for ECORR, EFAC, and EQUAD are assigned.
    Then a final timing model analysis is performed with the noise values fixed to the maximum likelihood values found through the noise analysis to determine any final changes to the timing model.
    Since these noise analyses only use the analytically marginalized timing model, one cannot determine whether the underlying timing model is changing in the presence of different noise combinations.
    Likewise, only using fixed noise is not necessarily representative of the noise level for variations in the timing model.
    Thus by replacing these final steps with a fully generalized timing model, one can simultaneously determine the noise and timing parameter distributions and track covariances.
    
    One can also use it to analyze each individual pulsar as they evolve over time in each of the timing parameters to discover how the interplay of noise and observation parameters influence the timing model and vice versa.
    One can use model comparison, to compare the evidence for different binary models, inclusion of time derivatives of binary parameters such as $\dot{P_{\mathrm{b}}}$, $\dot{\omega}$, and $\dot{x}$, and significance of less physical parameters like the frequency-dependent changes in the pulse-profile shape.
    Product space sampling combined with our generalized timing framework could also prove effective at determining whether there is evidence for the inclusion of intrinsic RN on top of the pulsar WN.
    
    Timing model parameters that are covariant with RN and GW signals can be directly included in full PTA searches to determine if the covariances are negatively affecting the GW search sensitivity, or mimicking a signal.
    This could be especially useful in searches for continuous GWs from individual super massive black hole binaries that have orbital periods similar to effects in the timing model like the binary period.

    Beyond the physical priors we placed upon several timing model parameters already in this work, one could apply informative prior distributions to each to better test the timing model response with theoretical distributions.
    For example, in our examination of J1640+2224 with the NANOGrav 5 yr data, we used an informative prior on the parallax parameter using the DM distance derived from \cite{Cordes2002}.
    Sometimes pulsar distances are very well known due to direct observations of companions, or high angular resolution observations \cite{Deller2019}.
    Since parallax is directly dependent on the pulsar distance, direct distance measurements allow for strong constraints on the parallax parameter.
    Parallax appears in timing residuals as a yearly, sinusoidal effect and is a large source of covariance with other power law and sinusoidal parameters at a period of a year.
    As of this paper, such constraints have never been used in tandem with a fully generalized timing model.
    These more informative priors can be extended to other parameters as well as parallax.
    For pulsars with well characterized companions, especially companion masses that have been directly observed like J1640+2224, tight constraints can be put on priors that govern companion mass.

\subsection{Conclusions}
    Throughout this work we have tested the methods of generalized Bayesian timing techniques with contemporaneous noise modeling and shown that they are broadly consistent with previous partially generalized methods as well as proven analyses like GLS timing and chi-squared gridding.
    However, we have shown generalized timing can improve upon the timing model constraints from these other methods.
    Furthermore, the discrepancies between our results and those of published timing models highlight that these models are always evolving and rely on solutions in often highly covariant spaces that may not always provide a stable and accurate solution.
    We have shown by combining a varying timing model with noise parameters, one can determine evidence for the inclusion of RN and determine the amount of covariance between different parameters in the presence of various models.
    In particular, we find evidence for the inclusion of intrinsic RN in PSR J2043+1711 for the first time.
    
    We find that using more physical and restrictive priors like on pulsar mass and parallax can cause the timing model parameters to change significantly as compared to other methods.
    We have placed the tightest constraints on the companion and pulsar masses for PSR J0740+6630 and showed a shift to even lower values than found in \cite{Fonseca2021}.
    
    Fully generalized timing, using the frameworks laid out in this paper, can be expanded and utilized in almost all aspects of pulsar timing: from determining individual pulsar timing solutions to full PTA GW analyses.
    These methods were developed with the ultimate goal of using timing parameters in combination with GW signals together in full Bayesian analyses to further improve detection capabilities of PTAs.

\acknowledgements

This work has been carried out by the NANOGrav collaboration, which is part of the International Pulsar Timing Array. 
ARK developed and implemented the code and all methodologies, and led the paper writing.
JSH helped implement and develop the code, provided much insight, and contributed to writing throughout the manuscript.
MAM provided much feedback on the project development and manuscript.
SRT, MV, and SJV developed much of the framework and background for the project, wrote initial code, and provided feedback throughout the project.
TC and EF provided feedback throughout the project and allowed the use of their data.
JS helped develop the DM Gaussian process code.
ZA, PTB, HB, PRB, MED, PBD, TD, JAE, RDF, ECF, WF, NGD, PAG, DCG, RJJ, MLJ, DLK, VMK, MK, MTL, DRL, JL, RSL, AM, JWM, NM, BWM, CN, DJN, TTP, BBPP, NSP, SMR, RS, BJS, IHS, KS, JKS, HMW, WWZ ran observations and developed the 12.5-yr data used.
EF, TC, TTP, PSR, AYK, SMR, PBD, IHS, ZA, LG, AP, MK, IC, PTB, HB, PRB, MD, TD, FAD, ECF, WF, NGD, DCG, RJ, MLJ, VMK, MTL, DRL, JL, AM, JWM, MAM, NM, BWM, AN, CN, DJN, NP, HAR, BS, CMT, SPT, JKS, HMW, WWZ ran observations and developed the data for \cite{Fonseca2021} used in the analysis of J0740+6620.

The NANOGrav project receives support from National Science Foundation (NSF) Physics Frontiers Center award number \#1430284 and \#2020265.
This work made use of the Super Computing System, Thorny Flat, at West Virginia University (WVU), which is funded in part by the  National Science Foundation (NSF) Major Research Instrumentation Program (MRI) Award \#1726534, and the Super Computing System Spruce Knob at WVU, which is funded in part by the National Science Foundation EPSCoR Research Infrastructure Improvement Cooperative Agreement \#1003907, the state of West Virginia (WVEPSCoR via the Higher Education Policy Commission) and WVU.
Portions of this work performed at NRL were supported by ONR 6.1 basic research funding.
TD and MTL are supported by an NSF Astronomy and Astrophysics grant (AAG) award No. 2009468.

\section*{Code and Data Availability}
\ark{Code used in this analysis is available on GitHub. Specifically, our implementations of \texttt{enterprise\_extensions} can be found \href{https://github.com/ark0015/enterprise_extensions/tree/ark-nltm-e_e-freeze}{\url{https://github.com/ark0015/enterprise_extensions/tree/ark-nltm-e_e-freeze}} and \texttt{Enterprise} at \href{https://github.com/ark0015/enterprise/tree/ark-nltm}{\url{https://github.com/ark0015/enterprise/tree/ark-nltm}}, and our analysis code can be found at \href{https://github.com/ark0015/enterprise_timing}{\url{https://github.com/ark0015/enterprise_timing}}.}
\ark{The chains, priors, parameters, jump proposal draws, timing parameters, and covariance matrix information for the analyses included in this work can be found on Zenodo at \href{https://doi.org/10.5281/zenodo.20330523}{\url{https://doi.org/10.5281/zenodo.20330523}}.}

\section*{Software} 
We use the software packages \texttt{Enterprise} \citep{Enterprise, Enterprise2024} and \texttt{enterprise\_extensions} \citep{enterpriseextensions} to setup our models, priors, and calculate the likelihoods. We use \texttt{PTMCMCSampler} \citep{PTMCMC} as the Markov Chain Monte Carlo sampler for our Bayesian analysis. We use \texttt{PINT} \citep{PINT} and \texttt{libstempo} \citep{libstempo}, which interfaces with TEMPO2 \cite{TEMPO2I,TEMPO2II,TEMPO2III}, to construct residuals from the timing model parameters. We also use, Astropy \citep{astropy2013,Astropy2018}, Matplotlib \citep{Matplotlib2007}, NumPy \citep{Numpy2020}, Python \citep{Python2007,Python2011}, and SciPy \citep{Scipy2020}.

\clearpage
\appendix

\section{Timing Model Changes}
\label{sec:appendixA}

\begin{table*}[!htb]
    \centering
    \begin{tabular}{@{} |c|ccc| @{}}
        \hline\hline
        Parameter & \multicolumn{3}{|c|}{PSR J2043+1711: NANOGrav 9-yr} \\
        & GLS: & Model B: & Model C: \\
         & No RN & No RN & No RN \\
        \hline
        $\lambda$ & $5.5653050058(3)$ & $5.5653050056(4)$ & $5.5653050056(3)$ \\
    	$\beta$ & $0.5927893202(5)$ & $0.59278932(6)$ & $0.59278932(6)$ \\
    	$\nu$ & $420.189443268538(2)$ & $420.189443268538(2)$ & $420.189443268538(2)$ \\
    	$\dot{\nu}$ & $-9.257(2)\times10^{-16}$ & $-9.257(2)\times10^{-16}$ & $-9.2591(4)(2)\times10^{-16}$ \\
    	$\mu_{\lambda}$ & $-8.97(7)$ & $-8.98(8)$ & $-8.97(8)$ \\
    	$\mu_{\beta}$ & $-8.5(1)$ & $-8.5(1)$ & $-8.5(1)$ \\
    	$\pi$ & $0.8(2)$ & $0.8(2)$ & $0.8(2)$ \\
    	$P_{\mathrm{b}}$ & $1.4822907865(1)$ & $1.4822907865(2)$ & $1.4822907865(2)$ \\
    	$T_\mathrm{asc}$ & $56174.30624072(1)$ & $56174.30624072(1)$ & $56174.30624072(1)$ \\
    	$x$ & $1.6239584(2)$ & $1.6239583(3)$ & $1.6239583(3)$ \\
    	$\epsilon_{1}$ & $-4.3(1)\times10^{-6}$ & $-4.3(2)\times10^{-6}$ & $-4.3(1)\times10^{-6}$ \\
    	$\epsilon_{2}$ & $-2.42(9)\times10^{-6}$ & $-2.4(1)\times10^{-6}$ & $-2.4(1)\times10^{-6}$ \\
        $\cos i$ & $0.12(1)$ & $0.12(2)$ & $0.12(2)$ \\
    	$m_{\mathrm{c}}$ & $0.17(1)$ & $0.18(2)$ & $0.18(2)$ \\
        $m_{\mathrm{p}}$ & $1.4(2)$ & $1.4(2)$ & $1.4(2)$ \\
        \hline
    \end{tabular}
    \caption{Median values of each timing model parameter posterior with its one sigma error for the NANOGrav 9-yr dataset.}
    \label{tab:J2043_tm_changes_pt1}
\end{table*}

\begin{table*}[!htb]
    \centering
    \begin{tabular}{@{} |c|ccc| @{}}
        \hline\hline
        Parameter & \multicolumn{3}{|c|}{PSR J2043+1711: NANOGrav 12.5-yr} \\
        & GLS: & Model B: & Model C: \\
         & No RN & Varied RN & Varied RN \\
        \hline
        $\lambda$ & $5.56530490141(9)$ & $\boldsymbol{\mathit{5.56530490118(6)}}$ & $\boldsymbol{\mathit{5.56530490118(8)}}$ \\
    	$\beta$ & $0.5927892344(1)$ & $\boldsymbol{\mathit{0.59278923406(1)}}$ & $0.5927892341(1)$ \\
    	$\nu$ & $420.189443214863(3)$ & $420.189443214862(1)$ & $\mathit{420.1894432148614(7)}$ \\
    	$\dot{\nu}$ & $-9.25931(9)(2)\times10^{-16}$ & $\mathit{-9.2593(6)\times10^{-16}}$ & $\mathit{-9.2589(4)\times10^{-16}}$ \\
    	$\mu_{\lambda}$ & $-8.815(2)$ & $\boldsymbol{\mathit{-8.824(1)}}$ & $\mathit{-8.827(9)}$ \\
    	$\mu_{\beta}$ & $-8.51(2)$ & $\mathbf{-8.5(1)}$ & $-8.5(2)$ \\
    	$\pi$ & $0.72(6)$ & $\mathbf{0.69(4)}$ & $0.69(6)$ \\
    	$P_{\mathrm{b}}$ & $1.48229078636(1)$ & $\mathbf{1.48229078637(1)}$ & $\mathbf{1.48229078637(1)}$ \\
    	$T_\mathrm{asc}$ & $56845.783966945(6)$ & $\mathbf{56845.783966942(5)}$ & $\mathbf{56845.78396694(5)}$ \\
    	$x$ & $1.6239581(2)$ & $\mathbf{1.6239581(1)}$ & $\mathbf{1.6239582(1)}$ \\
    	$\epsilon_{1}$ & $-4.23(7)\times10^{-6}$ &$\mathbf{-4.21(5)\times10^{-6}}$ & $\mathbf{-4.2(6)\times10^{-6}}$ \\
    	$\epsilon_{2}$ & $-2.68(4)\times10^{-6}$ & $\mathbf{-2.69(2)\times10^{-6}}$ & $\mathbf{-2.68(3)\times10^{-6}}$ \\
        $\cos i$ & $0.14(1)$ & $\mathbf{0.138(9)}$ & $\mathbf{0.135(9)}$ \\
    	$m_{\mathrm{c}}$ & $0.19(1)$ & $\mathbf{0.189(8)}$ & $\mathbf{0.185(9)}$ \\
        $m_{\mathrm{p}}$ & $1.6(2)$ & $\mathbf{1.6(1)}$ & $\mathbf{1.5(1)}$ \\
        \hline
    \end{tabular}
    \caption{Median values of each timing model parameter posterior with its one sigma error for the NANOGrav 12.5-yr dataset; the bold values denote when the posterior value is more constrained then the initial GLS values ($\sigma_{\mathrm{post}}<\sigma_{\mathrm{GLS}}$), and italic values indicate that the median value is not encompassed by the errors on the GLS best fit values.}
    \label{tab:J2043_tm_changes_pt2}
\end{table*}

\begin{table*}[!htb]
    \centering
    \begin{tabular}{@{} |c|ccc| @{}}
        \hline\hline
            Parameter & \multicolumn{3}{|c|}{PSR J1600-3053: NANOGrav 12.5-yr} \\
            & GLS: & Model B: & Model C: \\
             & No RN & No RN & No RN \\
        \hline
        $\lambda$ & $4.2646714996(1)$ & $\mathbf{4.26467149957(1)}$ & $\mathbf{4.2646714996(1)}$ \\
    	$\beta$ & $-0.1757867787(6)$ & $\mathbf{-0.1757867792(6)}$ & $\mathbf{-0.1757867793(6)}$ \\
    	$\nu$ & $277.9377112254727(2)$ & $277.9377112254727(2)$ & $277.9377112254727(2)$ \\
    	$\dot{\nu}$ & $-7.33863(4)\times10^{-16}$ & $\mathbf{-7.33862(4)\times10^{-16}}$ & $-7.33863(4)\times10^{-16}$ \\
    	$\mu_{\lambda}$ & $0.464(9)$ & $\mathbf{0.464(9)}$ & $\mathbf{0.463(9)}$ \\
    	$\mu_{\beta}$ & $-7.06(5)$ & $\mathbf{-7.06(5)}$ & $\mathbf{-7.06(5)}$ \\
    	$\pi$ & $0.51(8)$ & $\mathbf{0.51(8)}$ & $0.52(8)$ \\
    	$P_{\mathrm{b}}$ & $14.3484575509(2)$ & $14.3484575509(2)$ & $14.3484575509(2)$ \\
    	$T_{0}$ & $56165.2315(5)$ & $\mathbf{56165.2316(4)}$ & $56165.2316(6)$ \\
    	$x$ & $8.8016528(8)$ & $\mathbf{8.8016526(7)}$ & $8.801652(1)$ \\
        $\dot{x}$ & $-4(4)\times10^{-15}$ & $\mathbf{-4.1(5)\times10^{-15}}$ & $\mathbf{-4.1(6)\times10^{-15}}$ \\
    	$\omega$ & $181.86(1)$ & $181.86(1)$ & $181.86(1)$ \\
        $e$ & $0.000173728(9)$ & $\mathbf{0.000173729(9)}$ & $\mathbf{0.000173729(9)}$ \\
    	$\cos i$ & $0.43(8)$ & $\mathbf{0.45(5)}$ & $\mathbf{0.47(7)}$ \\
    	$m_{\mathrm{c}}$ & $0.30(9)$ & $\mathbf{0.32(9)}$ & $.3(1)$ \\ 
        $m_{\mathrm{p}}$ & $2(1)$ & $\mathbf{2.3^{+0.9}_{-0.7}}$ & $\mathbf{2.4^{+1.3}_{-0.8}}$ \\
        \hline
    \end{tabular}
    \caption{Median values of each timing model parameter posterior with its one sigma error for the NANOGrav 12.5-yr dataset; the bold values denote when the posterior value is more constrained then the initial GLS values ($\sigma_{\mathrm{post}}<\sigma_{\mathrm{GLS}}$), and italic values indicate that the median value is not encompassed by the errors on the GLS best fit values.}
    \label{tab:J1600_tm_changes}
\end{table*}

\begin{table*}[!htb]
    \centering
    \begin{tabular}{@{} |c|ccc| @{}}
        \hline\hline
            Parameter & \multicolumn{3}{|c|}{PSR J0740+6620: NANOGrav 12.5-yr} \\
            & GLS: & Model B: & Model C: \\
             & No RN & No RN & No RN \\
        \hline
        $\lambda$ & $1.8109385671(4)$ & $\mathit{1.8109385672(4)}$ & $1.8109385672(4)$ \\
        $\beta$ & $0.769733684(4)$ & $\mathbf{0.7697336845(4)}$ & $0.7697336845(4)$ \\
        $\nu$ & $346.531996527869(1)$ & $346.531996527869(1)$ & $346.531996527869(1)$ \\
        $\dot{\nu}$ & $-1.46374(8)\times10^{-15}$ & $-1.46373(8)\times10^{-15}$ & $-1.46373(8)\times10^{-15}$ \\
        $\mu_{\lambda}$ & $-2.78(7)$ & $-2.8(7)$ & $-2.8(7)$ \\
        $\mu_{\beta}$ & $-32.53(8)$ & $-32.53(8)$ & $-32.53(8)$ \\
        $\pi$ & $0.8(4)$ & $\mathbf{0.8(4)}$ & $0.8(4)$ \\
        $P_{\mathrm{b}}$ & $4.766944619(4)$ & $\mathbf{4.7669446189(4)}$ & $4.766944619(4)$ \\
        $T_\mathrm{asc}$ & $57275.60045624(3)$ & $57275.60045624(3)$ & $57275.60045624(3)$ \\
        $x$ & $3.9775561(4)$ & $3.977556(4)$ & $3.977556(4)$ \\
        $\epsilon_{1}$ & $-5.63(9)\times10^{-6}$ & $\mathbf{-5.64(8)\times10^{-6}}$ & $-5.65(9)\times10^{-6}$ \\
        $\epsilon_{2}$ & $-1.92(4)\times10^{-6}$ & $-1.92(4)\times10^{-6}$ & $-1.92(5)\times10^{-6}$ \\
        $\cos i$ & $0.02(6)$ & $\mathbf{0.04(4)}$ & $\mathbf{0.04(4)}$ \\
        $m_{\mathrm{c}}$ & $0.25(2)$ & $0.26(2)$ & $0.26(2)$ \\ 
        $m_{\mathrm{p}}$ & $2.0(3)$ & $\mathbf{2.1^{+0.3}_{-0.2}}$ & $\mathbf{2.1^{+0.3}_{-0.2}}$ \\
        \hline
    \end{tabular}
    \caption{Median values of each timing model parameter posterior with its one sigma error for the NANOGrav 12.5-yr dataset; the bold values denote when the posterior value is more constrained then the initial GLS values ($\sigma_{\mathrm{post}}<\sigma_{\mathrm{GLS}}$), and italic values indicate that the median value is not encompassed by the errors on the GLS best fit values.}
    \label{tab:J0740_tm_changes_pt1}
\end{table*}

\begin{table*}[!htb]
    \centering
    \begin{tabular}{@{} |c|ccc| @{}}
        \hline\hline
            Parameter & \multicolumn{3}{|c|}{PSR J0740+6620: \cite{Cromartie2020}} \\
            & GLS: & Model B: & Model C: \\
             & No RN & No RN & No RN \\
        \hline
        $\lambda$ & $1.8109385534(3)$ & $1.8109385534(3)$ & $1.8109385534(3)$ \\
        $\beta$ & $0.7697335659(3)$ & $\boldsymbol{\mathit{0.7697335664(3)}}$ & $\mathit{0.7697335664(3)}$ \\
        $\nu$ & $346.5319964932128(6)$ & $346.5319964932128(6)$ & $346.5319964932128(6)$ \\
        $\dot{\nu}$ & $-1.46389(2)\times10^{-15}$ & $\mathbf{-1.46388(2)\times10^{-15}}$ & $-1.46388(2)\times10^{-15}$ \\
        $\mu_{\lambda}$ & $-2.75(3)$ & $-2.75(3)$ & $-2.75(3)$ \\
        $\mu_{\beta}$ & $-32.43(4)$ & $-32.44(4)$ & $-32.44(4)$ \\
        $\pi$ & $0.5(3)$ & $\mathbf{0.6(3)}$ & $\mathbf{0.6(3)}$ \\
        $P_{\mathrm{b}}$ & $4.7669446191(1)$ & $4.7669446191(1)$ & $4.7669446191(1)$ \\
        $T_\mathrm{asc}$ & $57552.08324415(2)$ & $57552.08324415(2)$ & $57552.08324415(2)$ \\
        $x$ & $3.9775561(2)$ & $3.9775561(2)$ & $3.977556(2)$ \\
        $\epsilon_{1}$ & $-5.7(4)\times10^{-6}$ & $\mathbf{-5.71(4)\times10^{-6}}$ & $-5.71(4)\times10^{-6}$ \\
        $\epsilon_{2}$ & $-1.89(3)\times10^{-6}$ & $-1.89(3)\times10^{-6}$ & $-1.89(3)\times10^{-6}$ \\
        $\cos i$ & $0.045(4)$ & $\mathbf{0.046(4)}$ & $0.046(4)$ \\
        $m_{\mathrm{c}}$ & $0.258(8)$ & $\mathbf{0.259(8)}$ & $0.259(8)$ \\ 
        $m_{\mathrm{p}}$ & $2.14^{+0.20}_{-0.18}$ & $\mathbf{2.15^{+0.1}_{-0.09}}$ & $\mathbf{2.15^{+0.1}_{-0.09}}$ \\
        \hline
    \end{tabular}
    \caption{Median values of each timing model parameter posterior with its one sigma error for the \cite{Cromartie2020} dataset; the bold values denote when the posterior value is more constrained then the initial GLS values ($\sigma_{\mathrm{post}}<\sigma_{\mathrm{GLS}}$), and italic values indicate that the median value is not encompassed by the errors on the GLS best fit values.
    The median value and 68.3\% CI for the chi-squared gridding analysis of J0740+6620's mass by \cite{Cromartie2020} is $2.14^{+0.1}_{-0.09}$, which includes the median values of our model B and C analyses.}
    \label{tab:J0740_tm_changes_pt2}
\end{table*}

\begin{table*}[!htb]
    \centering
    \begin{tabular}{@{} |c|ccc| @{}}
        \hline\hline
            Parameter & \multicolumn{3}{|c|}{PSR J0740+6620: \cite{Fonseca2021}} \\
            & GLS: & Model B: & Model C: \\
            & No RN & No RN & No RN \\
        \hline
        $\lambda$ & $1.8109385406(2)$ & $1.8109385406(2)$ & $1.8109385406(2)$ \\
        $\beta$ & $0.7697334558(2)$ & $\mathit{0.7697334563(2)}$ & $\mathit{0.7697334563(2)}$ \\
        $\nu$ & $346.5319964608338(3)$ & $\mathbf{346.5319964608339(3)}$ & $346.5319964608338(3)$ \\
        $\dot{\nu}$ & $-1.46387(1)\times10^{-15}$ & $\mathbf{-1.463871(9)\times10^{-15}}$ & $-1.46387(1)\times10^{-15}$ \\
        $\mu_{\lambda}$ & $-2.74(1)$ & $\mathbf{-2.73(1)}$ & $-2.73(1)$ \\
        $\mu_{\beta}$ & $-32.48(2)$ & $-32.48(2)$ & $-32.48(2)$ \\
        $\pi$ & $1.0(2)$ & $\mathbf{0.9(1)}$ & $0.9(2)$ \\
        $P_{\mathrm{b}}$ & $4.76694461933(8)$ & $\mathbf{4.76694461932(8)}$ & $4.76694461932(8)$ \\
        $\dot{P_b}$ & $1(1)\times10^{-12}$ & $\mathbf{1.2(2)\times10^{-12}}$ & $\mathbf{1.2(2)\times10^{-12}}$ \\
        $T_\mathrm{asc}$ & $57804.73130889(2)$ & $57804.73130889(2)$ & $57804.73130889(2)$ \\
        $x$ & $3.97755608(1)$ & $\mathbf{3.97755607(9)}$ & $3.9775561(1)$ \\
        $\epsilon_{1}$ & $-5.68(3)\times10^{-6}$ & $\mathbf{-5.69(3)\times10^{-6}}$ & $\mathbf{-5.69(3)\times10^{-6}}$ \\
        $\epsilon_{2}$ & $-1.83(2)\times10^{-6}$ & $-1.83(2)\times10^{-6}$ & $-1.83(2)\times10^{-6}$ \\
        $\cos i$ & $0.043(3)$ & $\mathbf{0.043(3)}$ & $0.043(3)$ \\
        $m_{\mathrm{c}}$ & $0.251(5)$ & $\mathbf{0.251(5)}$ & $0.252(5)$ \\
        $m_{\mathrm{p}}$ & $2.05(7)$ & $\mathbf{2.06(6)}$ & $\mathbf{2.06^{+0.07}_{-0.06}}$ \\
        \hline
    \end{tabular}
    \caption{Median values of each timing model parameter posterior with its one sigma error for the \cite{Fonseca2021} dataset; the bold values denote when the posterior value is more constrained then the initial GLS values ($\sigma_{\mathrm{post}}<\sigma_{\mathrm{GLS}}$), and italic values indicate that the median value is not encompassed by the errors on the GLS best fit values.
    The median value and 68.3\% CI for the chi-squared gridding analysis with DMX bins of 3 of J0740+6620's mass by \cite{Fonseca2021} is $2.08(7)$, which includes the median values of our model B and C analyses.}
    \label{tab:J0740_tm_changes_pt3}
\end{table*}

\begin{table*}[!htb]
    \centering
    \begin{tabular}{@{} |c|ccc| @{}}
        \hline\hline
            Parameter & \multicolumn{3}{|c|}{PSR J1640+2224:NANOGrav 5-yr} \\
            & GLS: & Model B: & Model C: \\
        \hline
        RA & $4.364540752(4)$ & $\mathit{4.36454074(1)}$ & $\mathit{4.36454074(1)}$ \\
    	DEC & $0.390997111(3)$ & $0.39099711(1)$ & $0.39099711(1)$ \\
        $\nu$ & $316.12398431362(2)$ & $\mathit{316.12398431367(5)}$ & $\mathit{316.12398431367(6)}$ \\
    	$\dot{\nu}$ & $-2.812(1)\times10^{-16}$ & $\mathit{-2.815(3)\times10^{-16}}$ & $\mathit{-2.814(3)\times10^{-16}}$ \\
    	$\mu_{\mathrm{RA}}$ & $2.2(1)$ & $\mathit{2.6(4)}$ & $\mathit{2.5(4)}$ \\
    	$\mu_{\mathrm{DEC}}$ & $-11.0(1)$ & $-11.0(3)$ & $-11.0(3)$ \\
    	$\pi$ & $1.0(2)$ & $0.9(2)$ & $0.9(2)$ \\
    	$P_{\mathrm{b}}$ & $175.46066225(6)$ & $\mathit{175.4606622(2)}$ & $175.4606622(2)$ \\
    	$T_{0}$ & $51626.1796(4)$ & $\mathit{51626.1811(1)}$ & $\mathit{51626.181(1)}$ \\
    	$x$ & $55.3297183(7)$ & $55.329719(2)$ & $55.329719(2)$ \\
    	$\omega$ & $50.733(8)$ & $\mathit{50.736(2)}$ & $\mathit{50.736(2)}$ \\
        $e$ & $0.00079712(2)$ & $\mathit{0.00079725(4)}$ & $\mathit{0.00079727(5)}$ \\
    	$\cos i$ & $0(7)$ & $\mathbf{0.5(3)}$ & $\mathbf{0.7(4)}$ \\
    	$m_{\mathrm{c}}$ & $0.25(3)$ & $\mathit{0.3(2)}$ & $\mathit{0.4(8)}$ \\ 
        $m_{\mathrm{p}}$ & $1.4(3)$ & $1.1^{+1.0}_{-0.8}$ & $1.0^{+1.0}_{-0.8}$ \\
        \hline
    \end{tabular}
    \caption{Median values of each timing model parameter posterior with its one sigma error for the NANOGrav 5-yr dataset; the bold values denote when the posterior value is more constrained then the initial GLS values ($\sigma_{\mathrm{post}}<\sigma_{\mathrm{GLS}}$), and italic values indicate that the median value is not encompassed by the errors on the GLS best fit values.}
    \label{tab:J1640_tm_changes_5yr}
\end{table*}

\begin{table*}[!htb]
    \centering
    \begin{tabular}{@{} |c|ccc| @{}}
        \hline\hline
            Parameter & \multicolumn{3}{|c|}{PSR J1640+2224:NANOGrav 9-yr} \\
            & GLS: & Model B: & Model C: \\
        \hline
        $\lambda$ & $4.2584129337(3)$ & $\mathit{4.2584129325(3)}$ & $\mathit{4.2584129325(3)}$ \\
    	$\beta$ & $0.7689662382(3)$ & $\mathbf{0.768966238(3)}$ & $0.768966238(3)$ \\
        $\nu$ & $316.1239842341273(5)$ & $\mathit{316.1239842341285(6)}$ & $\mathit{316.1239842341285(5)}$ \\
    	$\dot{\nu}$ & $-2.81566(7)\times10^{-16}$ & $-2.81562(8)\times10^{-16}$ & $-2.81562(8)\times10^{-16}$ \\
    	$\mu_{\lambda}$ & $4.19(1)$ & $4.19(1)$ & $4.18(1)$ \\
    	$\mu_{\beta}$ & $-10.73(2)$ & $-10.72(2)$ & $-10.72(2)$ \\
    	$\pi$ & $-1.0(6)$ & $\boldsymbol{\mathit{0.2(3)}}$ & $\boldsymbol{\mathit{0.2(3)}}$ \\
    	$P_{\mathrm{b}}$ & $175.4606(1)$ & $175.4606(2)$ & $175.4606(2)$ \\
    	$T_{0}$ & $54784.4708(3)$ & $\mathbf{54784.4711(2)}$ & $\mathbf{54784.4711(2)}$ \\
    	$x$ & $55.329718(2)$ & $\mathbf{55.3297192(9)}$ & $\mathbf{55.3297192(9)}$ \\
    	$\dot{x}$ & $0.0(1)$ & $\mathbf{1.4(1)\times10^{-14}}$ & $\mathbf{1.4(1)\times10^{-14}}$ \\
    	$\omega$ & $50.7316(7)$ & $\mathbf{50.7322(4)}$ & $\mathbf{50.7322(4)}$ \\
    	$\dot{\omega}$ & $-0.00026(6)$ & $-0.00026(7)$ & $-0.00026(7)$ \\
        $e$ & $0.00079725(1)$ & $\mathbf{0.000797261(6)}$ & $\mathbf{0.000797261(6)}$ \\
    	$\cos i$ & $0.46(1)$ & $\mathbf{0.39(6)}$ & $\mathbf{0.39(6)}$ \\
    	$m_{\mathrm{c}}$ & $0.5(2)$ & $\mathbf{0.36(9)}$ & $\mathbf{0.37(9)}$ \\ 
        $m_{\mathrm{p}}$ & $3(3)$ & $\mathbf{2.1^{+0.6}_{-0.7}}$ & $\mathbf{2.2^{+0.6}_{-0.7}}$ \\
        \hline
    \end{tabular}
    \caption{Median values of each timing model parameter posterior with its one sigma error for the NANOGrav 9-yr dataset; the bold values denote when the posterior value is more constrained then the initial GLS values ($\sigma_{\mathrm{post}}<\sigma_{\mathrm{GLS}}$), and italic values indicate that the median value is not encompassed by the errors on the GLS best fit values.}
    \label{tab:J1640_tm_changes_9yr}
\end{table*}

\begin{table*}[!htb]
    \centering
    \scalebox{0.8}{
    \noindent\makebox[\textwidth]{
        \begin{tabular}{@{} |c|c|cc|cc| @{}}
            \hline\hline
                Parameter & \multicolumn{5}{|c|}{PSR J1640+2224:NANOGrav 12.5-yr} \\
                & GLS: & Model B: & Model C: & Model B: & Model C: \\
                & & Restricted Mass & Restricted Mass & Unrestricted Mass & Unrestricted Mass \\
            \hline
            $\lambda$ & $4.2584129903(2)$ & $\boldsymbol{\mathit{4.2584129892(2)}}$ & $\mathit{4.2584129892(3)}$ & $\mathit{4.2584129893(2)}$ & $\mathit{4.2584129891(3)}$ \\
        	$\beta$ & $0.7689661392(3)$ & $\boldsymbol{\mathit{0.7689661387(2)}}$ & $\mathit{0.7689661387(3)}$ & $\boldsymbol{\mathit{0.7689661388(1)}}$ & $\mathit{0.7689661388(3)}$ \\
            $\nu$ & $316.1239842170007(1)$ & $\mathbf{316.1239842170008(1)}$ & $316.1239842170008(2)$ &  $\mathbf{316.1239842170008(1)}$ & $316.1239842170008(2)$ \\
        	$\dot{\nu}$ & $-2.81539(3)\times10^{-16}$ & $\mathbf{-2.81539(2)\times10^{-16}}$ & $-2.81539(3)\times10^{-16}$ & $\mathbf{-2.81539(1)\times10^{-16}}$ & $-2.81539(3)\times10^{-16}$ \\
        	$\mu_{\lambda}$ & $4.191(8)$ & $\mathbf{4.194(5)}$ & $4.192(8)$ & $\mathbf{4.195(8)}$ & $4.193(8)$ \\
        	$\mu_{\beta}$ & $-10.73(1)$ & $\mathbf{-10.735(7)}$ & $-10.73(1)$ & $\mathbf{-10.735(6)}$ & $-10.73(1)$ \\
        	$\pi$ & $0.8(3)$ & $\mathbf{0.8(2)}$ & $0.8(3)$ & $\mathbf{0.9(1)}$ & $0.8(3)$ \\
        	$P_{\mathrm{b}}$ & $175.460661901(4)$ & $\mathbf{175.460661901(3)}$ & $175.460661901(5)$ & $\mathbf{175.460661902(3)}$ & $175.460661901(5)$ \\
        	$T_{0}$ & $55661.7742(5)$ & $\mathbf{55661.7743(1)}$ & $\mathbf{55661.7743(1)}$ & $\mathbf{55661.7739(1)}$ & $\mathbf{55661.774(3)}$ \\
        	$x$ & $55.329721(5)$ & $\mathbf{55.3297203(5)}$ & $\mathbf{55.3297203(8)}$ & $\mathbf{55.3297175(8)}$ & $\mathbf{55.329718(2)}$ \\
        	$\dot{x}$ & $0.0(7)$ & $\mathbf{1.16(5)\times10^{-14}}$ & $\mathbf{1.15(7)\times10^{-14}}$ & $\mathbf{1.16(4)\times10^{-14}}$ & $\mathbf{1.13(7)\times10^{-14}}$ \\
        	$\omega$ & $50.7317(1)$ & $\mathbf{50.7319(2)}$ & $\mathbf{50.7319(3)}$ & $\mathbf{50.7312(2)}$ & $\mathbf{50.7313(6)}$ \\
            $e$ & $0.00079726(2)$ & $\mathbf{0.000797261(3)}$ & $\mathbf{0.000797261(5)}$ & $\mathbf{0.000797249(4)}$ & $\mathbf{0.000797251(1)}$ \\
        	$\cos i$ & $0.5(2)$ & $\mathbf{0.43(4)}$ & $\mathbf{0.43(6)}$ & $\mathbf{0.58(3)}$ & $\mathbf{0.6(1)}$ \\
        	$m_{\mathrm{c}}$ & $0.2(8)$ & $\mathbf{0.38(6)}$ & $\mathbf{0.37(9)}$ & $\mathbf{0.7(1)}$ & $\mathbf{0.6(4)}$ \\ 
            $m_{\mathrm{p}}$ & $0.5(9)$ & $\boldsymbol{\mathit{2.2(5)}}$ & $\boldsymbol{\mathit{2.2^{+0.5}_{-0.7}}}$ & $\mathit{5.3^{+1.0}_{-0.9}}$ & $\mathit{5^{+3}_{-2}}$ \\
            \hline
        \end{tabular}
    }}
    \caption{Median values of each timing model parameter posterior with its one sigma error for the NANOGrav 12.5-yr dataset; the bold values denote when the posterior value is more constrained then the initial GLS values ($\sigma_{\mathrm{post}}<\sigma_{\mathrm{GLS}}$), and italic values indicate that the median value is not encompassed by the errors on the GLS best fit values.}
    \label{tab:J1640_tm_changes_12p5yr}
\end{table*}

\begin{table*}
    \label{tab:res_comp}
    \centering
    \begin{tabular}{|c|c|c|cc|cc|}
        \hline\hline
        & & \multicolumn{5}{c|}{Best-Fit Res.\ rms ($\mu\mathrm{s}$)} \\
        & & GLS: & \multicolumn{2}{c|}{Model B:} & \multicolumn{2}{c|}{Model C:} \\
        Pulsar & Dataset & No RN & No RN & Varied RN & No RN & Varied RN \\
    	\hline
    	J2043+1711 & \cite{9yr_timing} & $1.555$ & $\mathbf{1.547}$ & $\mathbf{1.548}$ & $1.555$ & $1.555$ \\
        & \cite{12p5yr_timing} & $2.990$ & $\mathbf{2.985}$ & $\mathbf{2.986}$ & $\mathbf{2.989}$ & $2.993$ \\
        \hline
    	J1600--3053 & \cite{12p5yr_timing} & $1.951$ & $1.953$ & $1.958$ & $1.951$ & $1.952$ \\
    	\hline
        & \cite{12p5yr_timing} & $3.829$ & $\mathbf{3.826}$ & $\mathbf{3.826}$ & $\mathbf{3.826}$ & $\mathbf{3.826}$ \\
    	J0740+6620 & \cite{Cromartie2020} & $4.343$ & $\mathbf{4.340}$ & $\mathbf{4.340}$ & $\mathbf{4.341}$ & $\mathbf{4.341}$ \\
    	& \cite{Fonseca2021} & $5.537$ & $\mathbf{5.535}$ & $\mathbf{5.535}$ & $\mathbf{5.535}$ & $\mathbf{5.535}$ \\
    	\hline
    	J1640+2224 & \cite{Demorest2013} & $6.359$ & $\mathbf{6.273}$ & $6.409$ & $\mathbf{6.281}$ & $\mathbf{6.327}$ \\
        & \cite{9yr_timing} & $4.005$ & $\mathbf{3.988}$ & $\mathbf{3.989}$ & $\mathbf{3.994}$ & $\mathbf{3.994}$ \\
    	& \cite{12p5yr_timing} & $4.572$ & $\mathbf{4.054}$ & $\mathbf{4.053}$ & $\mathbf{4.058}$ & $\mathbf{4.058}$ \\
    	\hline
    \end{tabular}
    \caption{Comparison of the root-mean-squared residuals when using the median values of each timing model posteriors in comparison to the best fit values of the GLS analysis corresponding to the denoted published dataset. 
    \cite{Demorest2013} corresponds to the NANOGrav 5-yr dataset, \cite{9yr_timing} the 9-yr dataset, and \cite{12p5yr_timing} the 12.5-yr dataset.
    Bold values indicate and improvement on the GLS rms residuals.}
\end{table*}

\clearpage
\section{Analytical vs Numerical Marginalization}
\label{sec:appendixB}

Here we investigate further any discrepancies between the recovered posteriors of the analytically--marginalized and the numerically--marginalized posteriors. We go into great depth of how we construct the numerically--marginalized posteriors in section \S\ref{subsec:NonlinearTimingModels} and thus will not reiterate that discussion here. To extract the analytically--marginalized posteriors, we utilize functionality of \texttt{Enterprise} \citep{Enterprise, Enterprise2024} to retrieve the realizations of the timing model for particular states of the likelihood function.

We choose to focus on J1640+2224 as it shows the greatest difference between the fully numerically--marginalized (model B) and partially numerically-marginalized (model C) timing model posteriors (see figure \ref{fig:J1640_Keplerian_Plots}). 
\ark{We note the caveat however, that as discussed in Section \S\ref{subsec:J1640_12p5yr}, the numerically--marginalized analyses discussed here were only able to reach a Gelman-Rubin split R-hat statistic for the least converged variable of 6.0 and an autocorrelation length of $1.8\times10^{5}$ in the numerically--marginalized analyses, respectively, after over 1.6 million MCMC steps. Again we expect that with enough samples, the numerically--marginalized analyses would converge, but due to computational constraints we are unable to achieve full convergence here. Thus the numerically--marginalized results may not represent the true distribution of each parameter. The analytically--marginalized results however are fully converged according to our standards with a Gelman-Rubin split R-hat statistic $<1.1$ and an autocorrelation length of $48$ for $4.0\times10^{5}$ samples.}
First, we perform a purely analytically--marginalized (model A) analysis on J1640+2224 and use 500 samples of with the highest likelihood to ensure we use the parameter regions that maximize the likelihood function. We only select 500 samples  to reduce the computational cost of extracting the analytically--marginalized coefficients from the likelihood function. We then use the 500 samples as the realizations of the likelihood function and extract the relevant timing parameters for each realization by backing out the timing model design matrix for a particular likelihood value. These realizations provide a distribution around each of the timing model parameters that have the highest likelihood. Since these parameters also use an effective sampled space (see section \S\ref{subsec:global_model_params}), after re-scaling to physical values, the realization distributions are directly comparable to the sampled timing model posteriors of the numerically--marginalized model B.

As explained in section \S\ref{subsec:BayesianTiminginEnterprise}, the timing model parameters for a fully or partially analytically--marginalized model are marginalized by assuming a first-order Taylor-series approximation of the timing model by multiplying the design matrix with a set of basis coefficients that represent the deviations of the timing model parameters around the pre-fit parameters. The coefficients are given a Gaussian prior with infinite variance to effectively analytically-marginalize over the desired timing model parameters. When numerically marginalizing over the timing model parameters, this study uses a uniform prior bounded by $\mu_{\mathrm{GLS}} \pm 50\sigma_{\mathrm{GLS}}$ for each timing model parameter. The differences in priors, in addition to directly sampling in each included timing model parameter, both lead to different likelihood functions and result in different parameter posteriors.

\begin{figure*}[!htbp]
    \centering
    \includegraphics[width=\textwidth]{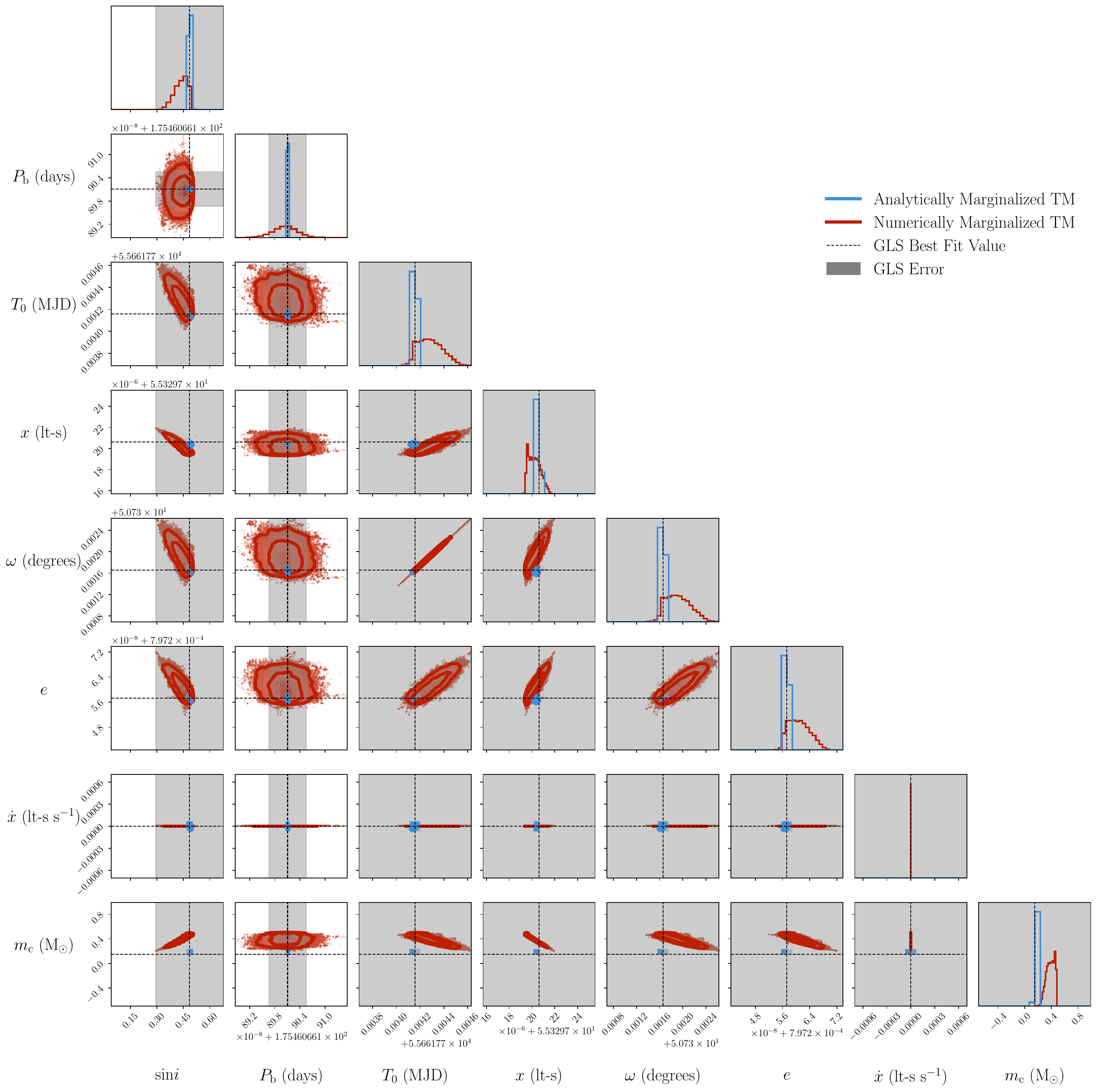}
    \caption{Comparison plots for PSR J1640+2224 using the NANOGrav 12.5-yr dataset \citep{12p5yr_timing} of the several binary parameters extracted analytically-marginalized timing model coefficients (model A; solid blue) and numerically-marginalized timing model samples (model B; solid red). Both models include white noise, but no red noise.
    }
    \label{fig:J1640_nltm_vs_ltm_coeffs_cornerplot}
\end{figure*}

In figure \ref{fig:J1640_nltm_vs_ltm_coeffs_cornerplot}, we show a comparison between the analytically--marginalized (model A) extracted coefficients and the numerically--marginalized (model B). As seen in the figure, the model A timing model coefficient realizations do not follow their initial Gaussian distribution prior with a mean of the GLS best fit value and infinite variance. The distributions for all analytically-marginalized timing model coefficients studied here closely sample around the GLS best fit value with little deviation. On the other hand, the bounded numerically--marginalized posteriors often shift away from the best fit value, albeit within the GLS error regions.

\begin{figure*}[!htbp]
    \centering
    \includegraphics[width=\columnwidth]{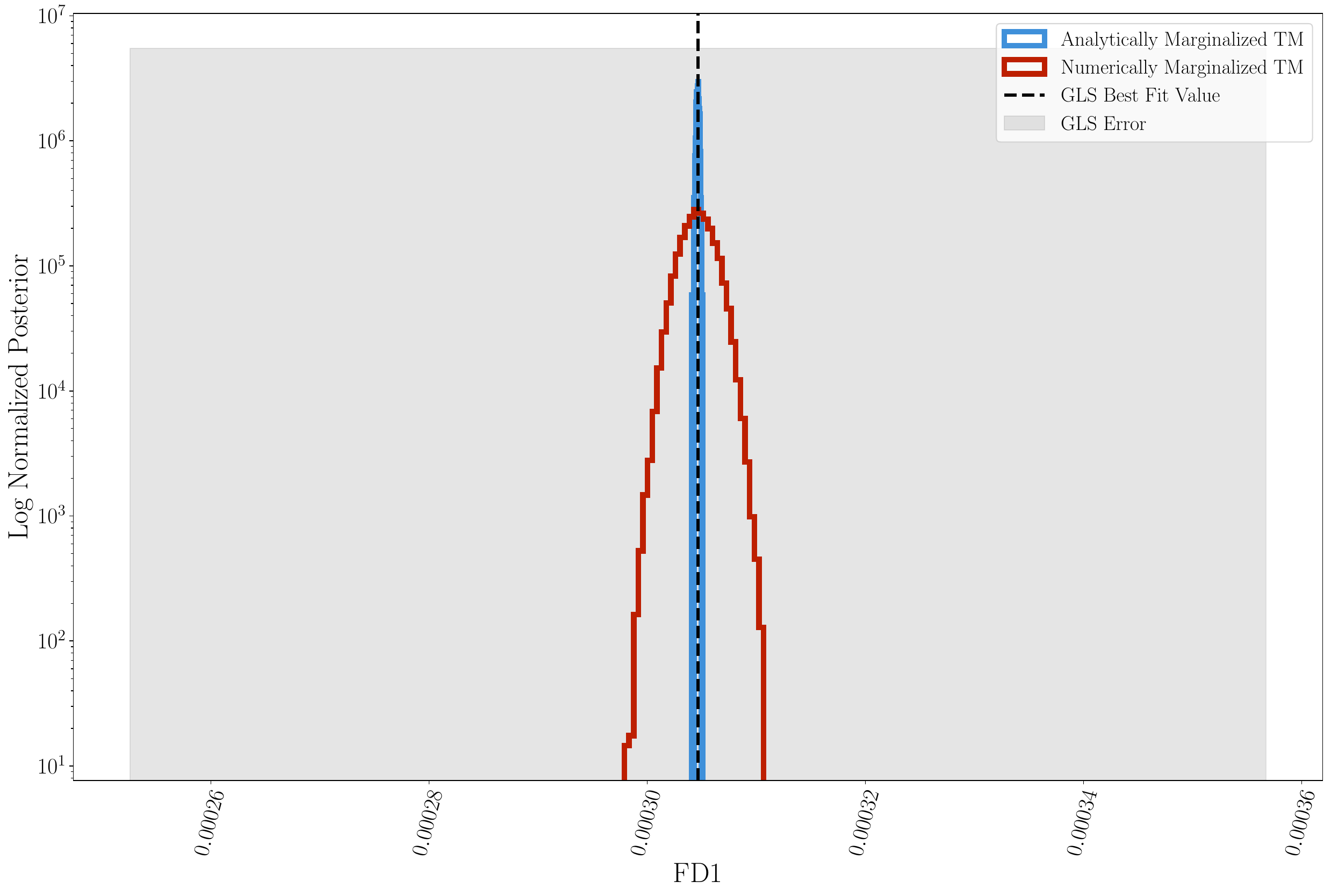}
    \caption{Comparison plots for the first FD parameter of PSR J1640+2224 using the NANOGrav 12.5-yr dataset \citep{12p5yr_timing} of extracted analytically-marginalized timing model coefficients (model A; solid blue) and numerically-marginalized timing model samples (model B; solid red). Both models include white noise, but no red noise.
    }
    \label{fig:J1640_nltm_vs_ltm_coeffs_FD1}
\end{figure*}

\begin{figure*}[!htbp]
    \centering
    \includegraphics[width=\columnwidth]{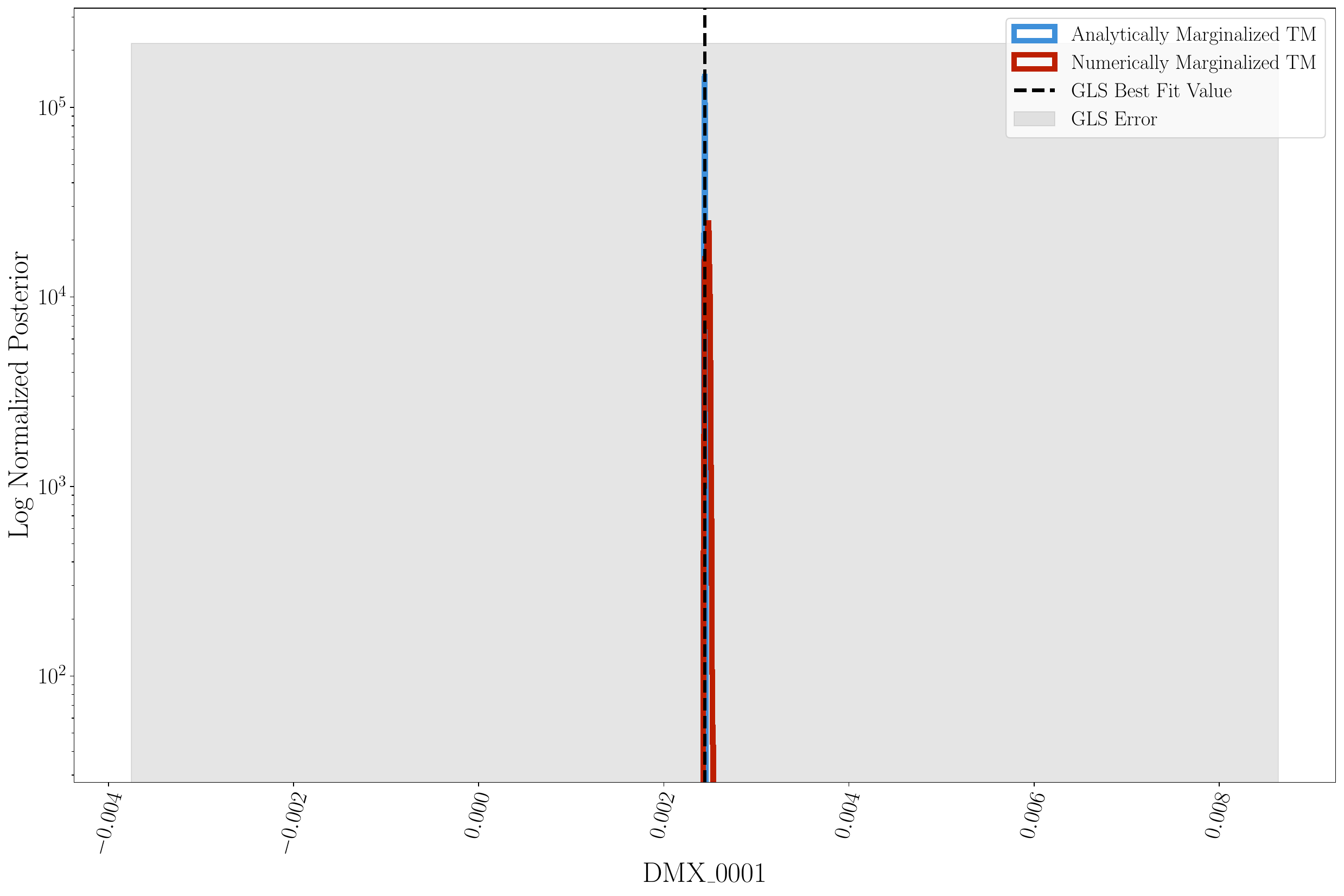}
    \caption{Comparison plot for the first DMX parameter for PSR J1640+2224 using the NANOGrav 12.5-yr dataset \citep{12p5yr_timing} of extracted analytically-marginalized timing model coefficients (model A; solid blue) and numerically-marginalized timing model samples (model B; solid red). Both models include white noise, but no red noise.
    }
    \label{fig:J1640_nltm_vs_ltm_coeffs_DMX_0001}
\end{figure*}

When investigating the ``nuisance'' parameters such as DMX and frequency dependent (FD) parameters, as shown in figures \ref{fig:J1640_nltm_vs_ltm_coeffs_DMX_0001} and \ref{fig:J1640_nltm_vs_ltm_coeffs_FD1}, we find little shift in the distribution for the numerically-marginalized posteriors, indicating the GLS best fit values are representative of the mean values for these parameters. This reinforces our findings that analytically marginalizing over these ``nuisance'' parameters does not greatly affect the non-``nuisance'' timing model parameters under the conditions of this study.

\clearpage
\bibliographystyle{aasjournal}
\bibliography{biblio.bib}

\end{document}